\documentclass[useAMS,usenatbib]{mn2e}

\usepackage{epsf,rotating}
\usepackage{amsmath}
\usepackage{subfigure}
\usepackage{amssymb}
\usepackage{color}

\newcommand{\lya}{Ly$\alpha$}

\newcommand{\hmpc}{h^{-1}{\rm Mpc}}

\newcommand{\cmc}{\;{\rm cm}^{-3}}

\newcommand{\msolar}{\;{\rm M}_{\odot}}

\newcommand{\gad}{{\sc Gadget-2}}

\newcommand{\ion}[2]{\hbox{#1\,{\sc #2}}}
\newcommand{\vw}{v_{\rm w}}

\newcommand{\aap}{A\&A}

\newcommand{\apj}{ApJ}

\newcommand{\mnras}{MNRAS}

\newcommand{\fesc}{f_{\mathrm{esc}}}

\newcommand{\rion}{{R_{\rm ion}}}
\newcommand{\rrec}{{R_{\rm rec}}}
\title[New EoR 21cm Simulations]{Simulating the 21-cm signal from reionisation including non-linear ionisations and inhomogeneous recombinations}

\author[S. Hassan et al.]{
\parbox[t]{\textwidth}{\vspace{-1cm}
Sultan Hassan$^1$, Romeel Dav\'e$^{1,2,3}$, Kristian Finlator$^{4,5}$, Mario G. Santos$^{1,6,7}$}
\\
\\$^1$ University of the Western Cape, Bellville, Cape Town, 7535, South Africa
\\$^2$ South African Astronomical Observatories, Observatory, Cape Town, 7925, South Africa
\\$^3$ African Institute for Mathematical Sciences, Muizenberg, Cape Town, 7945, South Africa
\\$^4$ Dark Institute of Cosmology, Niels Bohr Institute, University of Copenhagen, Copenhagen, 2100, Denmark
\\$^5$ New Mexico State University, Las Cruces, NM 88003, United States
\\$^6$ SKA SA, The Park, Park Road, Pinelands 7405, South Africa
\\$^7$ CENTRA, Instituto Superior T\'ecnico, Universidade de Lisboa, Lisboa 1049-001, Portugal
}

\begin{document}

\maketitle

 \begin{abstract}
We explore the impact of incorporating physically motivated ionisation
and recombination rates on the history and topology of cosmic
reionisation and the resulting 21-cm power spectrum, by incorporating
inputs from small-volume hydrodynamic simulations into our
semi-numerical code, {\sc SimFast21}, that evolves reionisation on
large scales.  We employ radiative hydrodynamic simulations to
parameterize the ionisation rate $\rion$ and recombination rate
$\rrec$ as functions of halo mass, overdensity and redshift.  We
find that $\rion$ scales super-linearly with halo mass ($\rion\propto
M_h^{1.41}$), in contrast to previous assumptions.  Implementing
these scalings into {\sc SimFast21}, we tune our one free parameter,
the escape fraction $f_{\rm esc}$, to simultaneously reproduce
recent observations of the Thomson optical depth, ionizing emissivity,
and volume-averaged neutral fraction by the end of reionisation.
This yields $f_{\rm esc}=4^{+7}_{-2}\%$ averaged over our $0.375\hmpc$ cells,
independent of halo mass or redshift, increasing to 6\% if we also
constrain to match the observed $z=7$ star formation rate function.
Introducing super-linear $\rion$ increases the duration of reionisation
and boosts small-scale 21-cm power by $\times 2-3$ at intermediate
phases of reionisation, while inhomogeneous recombinations reduce
ionised bubble sizes and suppress large-scale 21-cm power by $\times
2-3$.  Gas clumping on sub-cell scales has a minimal effect on the
21cm power.  Super-linear $\rion$ also significantly increases the
median halo mass scale for ionising photon output to $\sim 10^{10}
\rm M_\odot$, making the majority of reionising sources more
accessible to next-generation facilities.  These results highlight
the importance of accurately treating ionising sources and
recombinations for modeling reionisation and its 21-cm power spectrum.
\end{abstract}


\section{Introduction}

The epoch of reionisation (EoR) is the last global phase transition
of the Universe, during which the birth of first luminous sources
gradually ionised the intergalactic medium (IGM). Given that this
epoch remains mostly unexplored, current EoR studies are devoted
to answering its most basic questions: When did the EoR begin? What
are the sources responsible for driving reionisation? How did the
topology of reionisation evolve?  When did the EoR end?  Accurate
answers for these questions are crucial to understand the early
stages of galaxy formation and evolution.

Observations of the high-redshift quasars' \lya\ absorption
spectra~\citep{bec01,fan06} suggest that reionisation of the IGM
completed about $\mathrm{z} \sim 6$, though there is
some evidence for variations in this~\citep{pen14,bec15,char15}.
The recent cosmic microwave background (CMB) polarization measurements
by ~\citet{planck15} reported a Thomson electron scattering optical
depth out to the surface of last scattering of $0.066$, lower than
previous measures by WMAP~\citep{hin13}, thus reducing the need for
exotic ionising sources within high redshift star-forming
galaxies~\citep{rob15}.  These observations constrain the topology,
sources, and evolution of the EoR, albeit only crudely.  The detailed
evolution of spatially-inhomogeneous reionisation involves a complex
interplay between evolving source populations, the propagation of
ionising photons within a patchy IGM, and the enrichment of the
first galaxies and halos that may affect the nature of the ionising
sources.

A promising approach to tracking the evolution of neutral hydrogen
during the EoR is via its emission in the redshifted hyperfine 21-cm
line. The advantage of using the 21-cm line is that its brightness
temperature is directly proportional to neutral gas content
($\mathrm{x_{HI}}$) which makes it possible to study the three-dimensional
distribution of the neutral gas and the large scale structure during
the EoR~\citep{bar01,fur06}.  Hence the EoR is a key science goal
for current and future low-frequency ($\sim 150$~MHz) radio telescopes.
These experiments include the Low Frequency Array
(LOFAR)\footnote{http://www.lofar.org/}, the Murchison Widefield
Array (MWA)\footnote{http://www.haystack.mit.edu/ast/arrays/mwa/},
the Precision Array to Probe Epoch of Reionisation
(PAPER)\footnote{http://w.astro.berkeley.edu/~dbacker/eor/}, the
Hydrogen Epoch of Reionisation Array
(HERA)\footnote{http://reionisation.org} and eventually the Square
Kilometer Array (SKA-Low) \footnote{https://www.skatelescope.org}.
Owing to the relatively coarse angular resolution at these long
wavelengths, interpreting these observations fully requires
understanding how small-scale processes such as the production and
propagation
of ionising photons impact the large-scale neutral gas distribution.

To this end, numerous theoretical studies have focused on modeling
the EoR process and its expected 21-cm signal.  A key limiting
factor is the need for vast dynamic range that remains a computational
challenge.  Proper cosmological simulations of the EoR must resolve
the smallest proto-galaxies to track the local generation and
propagation of the ionizing photons, as well as their recombinations
which requires modeling the small-scale clumping of gas, from
sub-$\mathrm{kpc}$ scales up to the sizes of the largest ionising
bubbles during the late stages of the EoR, believed to be tens of
$\mathrm{Mpc}$~\citep[e.g.][]{sob14}. This requires both large-volume
simulations together with extremely high resolution.  For instance,
the forthcoming EoR surveys by SKA-Low will span a field-of-view
of about 10 degrees that corresponds to a size of $\sim 500$~Mpc
with $\sim 0.5$~Mpc resolution~\citep{ian15}, while the proto-galaxies
that are thought to provide the dominant source of ionising photons
likely have scale sizes below a kpc~\citep[e.g.][]{tilvi13}.

Early hydrodynamic simulations of the EoR applied post-processed
radiative transfer to cosmologically-evolved density
fields~\citep[e.g.][]{gne00,raz02,mel06,mcq07,thom09,ian14,bau15},
which enabled the study of reionisation topology but did not
self-consistently include the feedback effects of the ionisation
on galaxies.  More recent codes such as {\sc March}~\citep{fin09,fin13}
and {\sc Traphic}~\citep{pawlik08} have been developed to do full radiative
hydrodynamics in a cosmological galaxy formation code, including
feedback processes constrained to match available high-$z$
data~\citep{fin11}.  Unfortunately, these simulations are computationally
very expensive, and the requirement that the simulation resolve to
below the hydrogen cooling halo mass limit of $\sim 10^8 \rm M_\odot$
drives one to a quite small box size (see however~\citealt{so14,gne14}), typically $\la 10$~Mpc.

A complementary approach is to use a large volume ($\ga 100$~Mpc)
simulations that employ physically-motivated parameterizations to
determine the location and evolution of the source population, even
though they do not resolve individual galaxy sources.  These so-called
semi-numerical models have low computational cost since they do not
require radiative transfer and they determine the density field
evolution from analytic prescriptions such as excursion set
formalism~\citep{mesinger07,zah07,cho09,san10}.  A key free parameter
in such codes is the ionizing efficiency per unit (halo) mass,
generally assumed to be constant and tuned to match key observations.
Semi-numerical models have been shown to roughly reproduce the
ionisation history and 21-cm power spectrum as obtained from radiative
hydrodynamic simulations~\citep{zah11,maj14}.  However, the simplicity
of the source parameterisations and the fact that they neglect
recombinations means that they fail to account for all of the
relevant inhomogeneous physical processes required to properly
address the problem.

\citet{sob14} have recently improved their semi-numerical code 
by incorporating inhomogeneous recombinations using a sub-grid
prescription that self-consistently tracks the evolution of ionizing
sources and recombination systems, as well as accounting for feedback
effects. While an improvement over previous semi-numerical work,
the parameterisations are obtained without self-consistently accounting for 
galactic feedback processes, and hence still employ a constant
efficiency parameter (i.e. ionising photon rate per unit dark matter
halo mass $M_h$) to characterise the source population. In contrast,
radiative hydrodynamic simulations tuned to match high-$z$ galaxy
properties find that the star formation rate, and hence relatedly
the ionising photon rate, scales super-linearly with halo mass at
these epochs, typically as $M_{h}^{1.3-1.4}$~\citep{fin11}.
Furthermore, the clumping factor that controls the recombination
rate is set by a complex interplay between photon propagation as a
function of environment, which is best followed using full radiative
hydrodynamic simulations~\citep[e.g.][]{fin12}.
Hence there remains room for improvement in semi-numerical EoR models.

In this paper we aim to take the next step by implementing calibrated
relations for the non-linearly mass-dependent ionisation efficiency
and inhomogeneous recombination rates taken from high-resolution
radiative hydrodynamic (RT) simulations into the semi-numerical EoR code
{\sc SimFast21}~\citep{san10}.  We use simulations from \citet{fin15}
that have been previously calibrated to match various observations
of galaxies at high redshifts.  We derive new parametrizations for
the non-linear ionisation ($\rion$) and recombination ($\rrec$)
rates directly from this radiative hydrodynamic simulation as a
function of halo mass and redshift, complemented by a larger-volume
galaxy formation simulation without radiative transfer to help
bridge the small scales in the RT simulation to the large-scale
semi-numerical model.  We implement these new formulae for $\rion$
and $\rrec$ that track the evolution of the non-linear ionisations
and inhomogeneous recombinations into {\sc SimFast21}.  We constrain our one free model
parameter, the escape fraction, to simultaneously match observations
of the Planck Thomson optical depth, the late evolution of the Lyman
alpha mean opacity, and the ionising emissivity at $z\sim 5$.
Finally, we make predictions for the expected 21-cm power spectrum
from the EoR, and particularly investigate how our new physical
parameterisations alter the predicted signal. This new version of our
{\sc SimFast21} will soon be publicly available\footnote{https://github.com/mariogrs/Simfast21}.

This paper is organized as follows: In section \ref{sec:sims}, we
introduce the simulations used, namely {\sc SimFast21}, the radiative hydrodynamic
simulations, and the non-radiative larger-volume simulation.  In
section \ref{sec:param}, we present our new parametrizations of
$\rrec$ and $\rion$ from hydrodynamic simulations. We present our
key observables in section \ref{sec:results}, explore the impact of our new 
parameterisations in section \ref{sec:impact} and conclude in section \ref{sec:conc}.
Throughout this work, we adopt a $\Lambda$CDM cosmology in which
$\Omega_{\rm M}=0.3$, $\Omega_{\rm \Lambda}=0.7$, $h\equiv H_0/(100
\, \mathrm{km/s/Mpc})=0.7$, a primordial power spectrum index $n=0.96$, an
amplitude of the mass fluctuations scaled to $\sigma_8=0.8$, and
$\Omega_b=0.045$. We quote all results in comoving units, unless
otherwise stated.

\section{Simulations}\label{sec:sims}

We describe {\sc SimFast21}~\citep{san10} and, in particular, focus on how
the ionized regions are identified using a fixed efficiency parameter.
We then briefly describe the two complementary suites of the
state-of-the-art hydrodynamic simulations, namely a high-resolution
radiative hydrodynamic simulation~\citep[{\bf 6/256-RT};][]{fin15}, and
a larger-volume cosmological hydrodynamic simulation without radiative
transfer~\citep[{\bf 32/512};][]{dav13}, which we employ to obtain new
parametrizations of the non-linear ionisation and inhomogeneous recombination
rates. A summary of our simulations is shown below in Table~\ref{simtab}.

\subsection{\sc SimFast21}\label{sec:simfast21}

{\sc SimFast21} is a semi-numerical code that predicts the redshifted
21-cm signal from cosmic reionisation. {\sc SimFast21} simulation
uses a Monte-Carlo approach to evolve the density field from a
Gaussian random initial state to form collapsed structures based
on a spherical collapse density threshold. This prescription generally
follows the algorithm described in \citet{mesinger07}, which we
briefly review here.

At very high redshift, the dark matter density field is distributed linearly
onto a grid, and the linear gravitational corrections are added
by applying the \citet{zeldovich70} approximation to evolve to lower
redshifts.  The collapsed dark matter halos are specified using the
excursion-set formalism with an overdensity threshold of $\rm \delta_c(z)\sim
1.68/D(z)$, where $\rm D(z)$ is the linear growth factor.

Ionized regions are identified using a similar form of the
excursion-set algorithm, based on the assumption of a
constant efficiency parameter of ionising photon per unit halo mass, $\zeta$. 
The ionisation condition
for any given region is simply the amount of collapsed dark matter
halo $f_{coll}$ compared with the efficiency parameter $\zeta$.
In other words, the region is considered to be fully ionized if:
\begin{equation}
\centering
f_{coll}\geq \zeta^{-1}, 
\label{barrier}
\end{equation}
and fully neutral if not.  For single cells which are not covered by ionised bubbles, we 
set their ionised fraction to $f_{coll}\zeta$.  Using this condition, the ionisation field is generated 
that is the main input required to compute the predicted 21-cm signal.
$\zeta$ encapsulates a mixture of the ionisation and recombination
processes such as ionising radiation escape fraction and recombinations by
Lyman Limit Systems (LLS).  However, it implicitly assumes that the recombinations 
trace the halos in the same way as the ionisations, which
is unlikely to be true in detail.  Nonetheless, $\zeta$ can be tuned to match
observations and yield predicted 21-cm power spectra.

In \S\ref{sec:modify} we will describe our improvements to this scheme.  In particular, we
aim to improve the following aspects from the previous version of {\sc SimFast21}:
\begin{itemize}
\item No explicit modeling of recombinations.  Recombinations are only implicitly
modeled via the constant efficiency parameter, which does not account for inhomogeneities
in the density field and the local clumping factor.
\item The use of a constant efficiency parameter.  We will show that this is not 
an optimal description of the ionisation rate as a function of halo mass.
\item Regions identified as fully ionised are set to have a neutral fraction of zero.
More realistically, they should have a small neutral fraction appropriate for an
optically-thin portion of the Universe in ionisation equilibrium.
\item  The density field spatial distribution.  The original
code uses nearest grid point assignment of density field to cells, but this can result
in a biased density field distribution with unexpected voids when applying the \citet{zeldovich70}
approximation. 
\end{itemize} 

A typical run with {\sc SimFast21} will have a volume of several hundred
Mpc and a cell size of several hundred kpc.  Hence this method
cannot resolve the galaxies that are the sources of ionising photons
in the EoR, nor the clumped density field that governs recombinations.
Therefore, to develop a better sub-grid model for these processes,
we need to employ high-resolution hydrodynamic simulations of galaxy
formation that can resolve these processes directly.  Next
we describe these hydrodynamic simulations.

\subsection{6/256-RT simulation}

To resolve the smallest significantly star-forming halos, we use the recent
radiative transfer cosmological hydrodynamic simulation described
more fully in \citet{fin15}.  This simulation uses the {\sc Gadget-3}
Smoothed Particle Hydrodynamics (SPH) code~\citep{spr05}, merged
with the radiative transfer code {\sc March}~\citep{fin09,fin12}
to run a fully self-consistent radiative hydrodynamic simulation
of early galaxy formation.  Built on the version of {\sc Gadget}
developed by \citet{opp08}, this code incorporates well-constrained
models for star formation-driven feedback processes, that do a
generally good job of matching observed lower-redshift galaxy and
IGM properties~\citep{opp08,dav11a,dav11b,dav13}.

The particular run we use, which we will call {\bf 6/256-RT}, employs a
small volume of $6\hmpc$ and evolves $2\times256^3$ particles.
The mass of each gas particle is $2.3\times 10^5
\rm M_\odot$ whereas the dark matter particle is about $5.6\times$ more
massive.  We therefore can resolve halos down to hydrogen cooling
limit at $10^{8}\rm M_\odot$ with 65 (dark matter+gas) particles, and
galaxies down to stellar masses of $7.4\times 10^{6} \rm M_\odot$  with
32~gas particles.  The equivalent Plummer gravitational softening
length employed is 469~pc (comoving), corresponding to around 50~physical
pc at the epoch of interest.  It is crucial that our
simulation resolves halos down to the hydrogen cooling
limit, and therefore should be resolving essentially all the ionising
photon output during the EoR under the assumption that halos below
the hydrogen cooling limit do not contribute substantially to the
photon budget, as is now generally believed~\citep[e.g.][]{wis09}.

This simulation also includes a subgrid Monte Carlo model for kinetic
galactic outflows, following the ``ezw" prescription described in
\citet{dav13}.  The two free parameters are the mass loading factor
$\eta$ and wind speed $v_w$, which vary with galaxy velocity
dispersion $\sigma$ that is calculated using an on-the-fly
friends-of-friends galaxy identification code.  The ezw prescription
employs momentum-driven wind scalings ($v_w\propto\sigma$ and
$\eta\propto \sigma^{-1}$) in sizable galaxies ($\sigma>75\, \rm kms^{-1}$),
and energy driven scalings ($\eta\propto \sigma^{-2}$) in dwarf
galaxies.  To mock up outflows blowing channels through the interstellar medium (ISM),
hydrodynamic forces are turned off until either the particle reaches
10\% of SF critical density or a time of $1.95\times10^{10}/(\vw
(\rm kms^{-1}))$~yr has passed.  These scalings result in many predictions
for galaxy and IGM properties that agree reasonably well with
observations~\citep{som15}.  This version of {\sc Gadget}
also includes a model for chemical enrichment following Type~II and
Type~Ia supernovae and stellar evolution, as well as primordial and
metal cooling.  The star formation model follows the multi-phase
model in \citet{spr03b}, with a density threshold of $n_{\rm H}=0.13
\cmc$.  A \citet{cha03} initial mass function (IMF) is assumed
throughout.  Radiative cooling follows the prescription in \citet{kat96}
to include the primordial cooling, and \citet{sut93} collisional ionisation 
equilibrium tables to account for cooling from metal lines.

The radiative transfer is done during the simulation run on a grid
of $32^3$ voxels, and its evolution is tracked using a moment-based
method, together with a long characteristics code periodically
employed to calculate the Eddington tensor in each cell in order
to close the moment hierarchy~\citep{fin11}.  The non-equilibrium
ionisation state of the gas is followed by interpolating the
ionisation field to the location of each gas particle.  Sixteen
separate frequency groups are followed, evenly spaced between
$1-10$~Ryd.

While the code is high resolution by cosmological standards, it
lacks sufficient resolution to directly predict the escape of
ionising photons from star-forming regions.  Hence there is still
an assumption required for the escape fraction of ionising photons,
$\fesc$.  In this run, a mass- and redshift-dependent $\fesc$ is
employed, which results in $\fesc=0.8$ in halos with mass
$\rm M_h<10^8M_\odot$, dropping with both mass and time to around 5\%
at $z=5$ in more massive halos; the full formula is in equation~1
of \citet{fin15}.

\citet{fin15} showed that this simulation, with its $\fesc$ as
assumed, can simultaneously match two key EoR observations, namely
the optical depth to Thomson scattering from the Wilkinson Microwave
Anisotropy Probe~\citep[WMAP;][]{hin13} polarization-temperature
cross-correlation, and the volume-averaged neutral fraction at
$z\sim 6$~\citep{fan06} measured from the Ly$\alpha$ forest.  Hence
it is a plausible model for the evolution of the ionising background
during the EoR.

As an important aside, we point out an inconsistency in this paper
regarding our assumed $\fesc$:  We will use this RT simulation's
output to provide a description for the ionising photon rate that
we will implement into {\sc SimFast21}, but then we will re-tune
$\fesc$ in our semi-numerical runs to a substantially different
value than what was assumed in the RT simulation.  Ideally, we would
re-run our RT simulation with the new $\fesc$, and iterate until
convergence.  Beyond the prohibitive computational cost to do so,
there is a more fundamental reason why this is not feasible:  The
volume of our RT simulation is too small to fully capture even the
moderately massive halos that (as we show later) contribute
significantly to the ionising photon budget particularly during the
latter stages of the EoR.  Therefore, the $\fesc$ used
in the RT simulation is tuned to high values at early stages of the
EoR to compensate for those missing halos. However, the critical
input from RT simulation is the slope of the star formation rate --
halo mass (SFR$- M_{h}$) relation
(SFR~$M_{h}^{1.3-1.4}$), which is sensitive to feedback from stellar
energy and the UVB at low masses. Hence it is appropriate to boost
the $\fesc$ in the RT simulation in order to model a realistic UVB
evolution, as this in turn yields a more realistic SFR($M_{h}, z$)
prediction.  Although the $\fesc$ assumed in the RT simulation is
not self-consistent, the fact that it reproduces key EoR observables
including the UV luminosity function means that it likely represents
a plausible evolution for the ionising photon output and neutral
gas density field over the scales that are modeled. Hence while at
present there is no practical way to avoid this inconsistency, 
it nonetheless provides an appropriate characterisation of the parameters
we need for our large-scale modeling.

\subsection{32/512 simulation}

While the RT simulation described above accurately follows the
interplay between photoionisation and galaxy formation, it lacks
the volume to produce sufficiently large halos that are an important
contributor to reionisation.  Since we want to evolve large volumes
with {\sc SimFast21}, we would like to have an ``intermediate" sized
simulation that effectively bridges the gap, and allows the computation
of ionisation and recombination rates in more massive halos.  For
this purpose, we will use a larger-volume cosmological hydrodynamic
simulation that does not include radiative transfer.  As discussed
in \citet{fin12}, radiative transfer effects are only important in
``photo-sensitive" halos with $M_h\la 10^{9.3}M_\odot$, and hence
we will only utilize this simulation for halos larger than this.

The simulation we employ has been fully described in \citet{dav13};
here we review the basic details.  This simulation uses the
\gad~N-body+SPH code~\citep{spr05} to model a cubic volume of
$32\hmpc$ on a side and $512^3$ particles of dark matter and gas
each; we refer to this as our {\bf 32/512} run.  Each gas particle has a
mass of about $\rm 4.5\times 10^6 M_\odot$, allowing us to resolve
galaxies down to stellar masses of around $M_{\rm *,lim}=1.4\times
10^8 M_\odot$, and halo masses of around $\rm 10^9M_\odot$.

Most of the physics is identical to the RT run, including the
radiative cooling, star formation following \citet{spr03a}, the
chemical evolution model, and the ``ezw" prescription for galactic
outflows.  One difference is that, unlike in the RT run, ionisation
equilibrium is assumed throughout, taking a spatially-uniform
ionising background as given by \citet{haa01}.  Also, this simulation
includes a model to quench star formation in massive galaxies, but
this does not come into play at the early epochs that we consider
here.

The overall strategy is to parameterise the ionisation and recombination
rates taken from the {\bf 6/256-RT} and {\bf 32/512} runs, and insert those
parameterisations into {\sc SimFast21}.  Next we describe how this is
done.

\section{Parameterizing ionisations and recombinations}\label{sec:param}

We aim to obtain new parameterizations of $\rion$ and $\rrec$ that
capture environmental and feedback effects directly from the {\bf 32/512}
and {\bf 6/256-RT} simulations, in order to implement them into {\sc  SimFast21}.
To begin, we must determine the ionisation and recombination rates
of our simulated gas particles.

\subsection{Ionisation Rate, $\rion$}

To determine the ionisation rate of each gas particle, we take
directly from the {\bf 6/256-RT} and {\bf 32/512} simulations output the star formation
rate (SFR) and metallicity (Z).  We then compute the ionisation
rate using Equation~(2) in~\citet{fin11}, which is derived from
~\citet{sch03} models.  Given an ionisation photon output rate for
each particle, we then sum this over all the star-forming gas
particles within a given halo to determine $\rion$ for that halo.

\begin{figure}
\centering
\setlength{\epsfxsize}{0.5\textwidth}
\centerline{\includegraphics[scale=0.5]{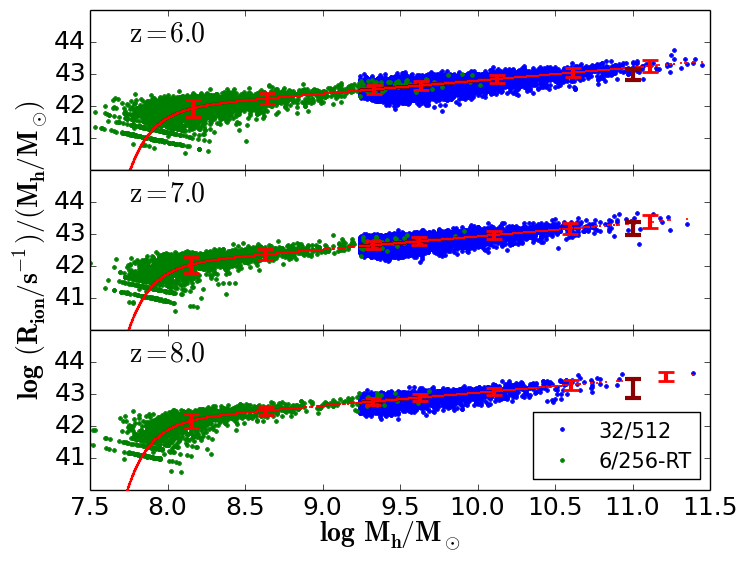}}
\caption{Ionization rate $\rion$ computed from {\bf 6/256-RT} (blue dots) and {\bf 32/512} (green) simulations. The overlap between {\bf 6/256-RT} and {\bf 32/512} simulations occurs in the halo mass range of $10^{9.3-10}\, \msolar$, and the two simulations yields similar results there. The red dots are computed using our fitting function, Equation~\eqref{eq:nion}. The red vertical error bars represent $\sigma$ values for ${\rm log\,M_{h}}$ bin size of $0.5$. The scatter is fairly small, so is unlikely to systematically impact the 21-power spectrum. The dark red vertical error bars at ${\rm 10^{11}\, M_{\odot}}$ are computed using SFR  measurements from \citet{bozi13}.  Our fitting function nicely reproduces the ionisation rate $\rion$ computed from our hydrodynamic simulations and only higher by a factor of $\times 1.5$ than \citet{bozi13} observations.}
\label{fig:nion}
\end{figure}

In Figure~\ref{fig:nion}, we show $\rion$ as a function of halo
mass for halos at $\rm z=6,7,8$ (top to bottom panels).  To increase
dynamic range, we combine the {\bf 6/256-RT} (green dots) and {\bf
32/512} (blue dots) simulations, thereby covering dark matter halo
masses of $M_h\sim 10^{8-11.5}\, \msolar$.  Importantly, the values
of $\rion$ among the two simulations is quite similar in the overlap
region of $M_h\sim 10^{9.3-10} M_\odot$, with only a very slight
tendency for the smaller volume to have higher $\rion$.  At low
halo masses ($M_h\la 10^8M_\odot$) we see a turn-down in $\rion$
owing to photo-suppression of galaxy formation~\citep{fin11}. 
At high masses, $\rion$
follows a power law that is approximately $\rion\propto M_h^{1.4}$,
consistent with what the dependence on SFR with $M_h$ found in
similar simulations by \citet{fin11}.

The red vertical error bars show the 1$\sigma$ scatter
in $\rion$ in ${\rm log\,M_{h}}$ bins of $0.5$.  At ${\rm M_{h} = 10^{10}
M_{\odot}}$, $\sigma \sim 0.12$ independent of redshift. Note that
when implementing $\rion$ into {\sc SimFast21}, we will average
over many halos in each cell which will reduce the scatter further,
therefore the scatter is unlikely to systematically impact the
21-power spectrum. Qualitatively, introducing scatter in $\rion$
would introduce larger HII bubbles in diffuse regions and smaller
ones in overdense ones, blurring the effect of the $\rion$ dependence
on $M_{h}$ and suppressing large-scale fluctuations.  These effects
are likely to be small given the small scatter, which only increases
substantially at the lowest masses that (we will show later) are
not driving reionisation.  The key goal of the current work is to
improve on previous work by allowing $\rion$ to vary with halo mass
in a physically-motivated way rather than assuming a constant value.
We therefore defer a detailed investigation into the impact of
scatter to future work.

To check if these simulated values for $\rion$ are
reasonable, we compare to the SFR inferred from observations using
abundance matching by \citet{bozi13}, shown as the dark red vertical
error bars at ${\rm 10^{11}\, M_{\odot}}$ (this is the lowest halo mass
at which they computed the SFR).  To convert from SFR to
$\rion$, we invert the process described above, assuming
a metallicity corresponding to the mean metallicity of star-forming
gas at that epoch.  The agreement is within the quoted uncertainty by 
\citet{bozi13}, but in detail
we see that our hydrodynamic simulations ({\bf
6/256-RT, 32/512 }) produce an $\rion$ that is higher by a factor
$\times 1.5$ compared to that inferred by \citet{bozi13}.
We will return to this point in \S\ref{sec:fesc}.

To implement $\rion$ into {\sc SimFast21}, we construct a fitting function
for $\rm \rion(M_h,z)$ whose mass dependence is analogous to a Schechter
function, namely a power-law on one end and an exponential cutoff
on the other, and whose redshift dependence is as a power law in
the scale factor\footnote{This is also taken from ~\citet{fra14}, where 
a similar formula used to parameterize the DLA cross-section; see their eq.~(4.4).}
, namely:
\begin{equation}\label{eq:nion}
\mathrm{\frac{\rion}{M_{h}} =  A(1 + z)^{D} ( M_{h}/B )^{C} \exp\left( -( B/M_{h})^{3.0} \right) } . 
\end{equation}
We perform a minimization to determine the best-fit parameters to be
$\mathrm{A} =1.08\times 10^{40}  \msolar^{-1} $s$^{-1}$, $\mathrm{B} = 9.51\times 10^{7}\msolar$, $\mathrm{C} = 0.41$ and $\mathrm{D} = 2.28$.  
Note that equation~\eqref{eq:nion} shows that $\rion$ scales as $\mathrm{M_{h}^{1.41}}$, 
which is consistent with the SFR$-\mathrm{M_{h}}$ relation that previously found by \citet{fin11}.

We note that at low redshifts (e.g. today), the high-mass end of
$\rion(M_h)$ would likely have a sharp turn-down owing to quenching
of star formation in $M_h\ga 10^{12}M_\odot$ halos~\citep[e.g.][]{gab12}.
We do not produce such large halos by $\rm z=6$ in these hydro simulations,
although it is possible that our {\sc
SimFast21} volume will be sufficiently large to yield halos above
this mass.  Nonetheless, there are suggestions that the quenching
mass scale is higher at high redshifts~\citep[e.g.][]{dek09,gab14}.
For instance, the best-fit parameterisation of the equilibrium galaxy
formation model in \citet{mit15} suggests
that at $z=6$, the quenching mass scale is close to $M_h\sim 10^{15}M_\odot$,
which is larger than any halo existing during the EoR.  Hence we
ignore the possibility of a high-mass turn-down in $\rion$ for now.

\subsection{Recombination Rate, $\rrec$}

Since the recombination rate $\rrec$ depends on the density field
and its clumping, we choose to parameterise $\rrec$ per unit volume
in terms of the local scaled density $\Delta\equiv \rho/\bar\rho$,
where $\rho$ is the matter density and $\bar\rho$ is the cosmic
mean at that redshift.  Then in {\sc SimFast21}, we can apply this
recombination rate to individual cells where we can compute the
scaled density.  Also, because we want to characterise the inhomogeneous
recombination rate during the EoR, we can only use the full radiative
hydrodynamic simulation for this, namely {\bf 6/256-RT}, and we do not
employ the {\bf 32/512} run here.  Unfortunately, this limits our dynamic
range and requires a larger extrapolation to cover the range of
overdensities that are achieved in our large {\sc SimFast21} volume,
which is an unavoidable limitation given computational capabilities.

Since the clumping and hence $\rrec$ is sensitive to
scale~\citep[e.g.][]{fin13}, we must choose a particular scale over which
to compute $\Delta$ and hence $\rrec$.  Here we choose to subdivide
our $6\hmpc$ simulation into 16 cells per side, which
yields a cell size of $0.535$~Mpc (comoving).  This then necessarily
fixes the cell size that we will use for our {\sc SimFast21} runs to a
value that enables us to feasibly simulate sufficiently large volumes
for computing the 21-cm power spectrum on the scales of relevance for
current and upcoming observations.  We smooth the mass of
each gas particle onto these cells using a 5th-order B-spline kernel
with 128 neighbours as utilised in the 6/256-RT simulation \citep{fin15}.

Within each so-defined cell in the {\bf 6/256-RT} simulation, we compute
the volume-averaged recombination rate density, $\mathrm{\rrec/V}$,
as follows:
\begin{equation}\label{rrec1}
\mathrm{\rrec/V = \alpha_{a} <  n_{HII}\, n_{e} >_{V}/x_{i} }\, ,
\end{equation}
where $\mathrm{n_{HII}}$ is ionized hydrogen number density,
$\mathrm{n_{e}}$ is the free electrons number density, $\mathrm{x_{i}}$ 
is the cell's ionised volume fraction, and  $\alpha_{a}
= 4.2 \times 10^{−13} \mathrm{cm^{3}\, s^{-1}}$, that corresponds
to a temperature of $10^{4}\, \mathrm{K}$.  We consider case-A
recombination following the recent paper by \citet{kaur14}, where
it has been shown that case-A recombination photons mimic the
scenario of those photons redshifted out of resonance, and therefore
can no longer be ionising photons. We compute
equation~\eqref{rrec1}
only in cells above a specific ionisation threshold $\mathrm{x_{i}}>0.95$.
 This is because in our model, {\sc SimFast21} 
assumes that ionisations must overcome recombinations in fully ionized medium,
once the ionisation condition is statisfied (eq.~\ref{barrier1}). 
Therefore, we want to make sure the recombinations rates computed from {\bf 6/256-RT} correspond to similarly
ionised regions.  Then we divide by $\mathrm{x_{i}}$ 
in equation~\eqref{rrec1} to correct for the residual neutral fraction. 

\begin{figure}
\setlength{\epsfxsize}{0.5\textwidth}
\centerline{\includegraphics[scale=0.48]{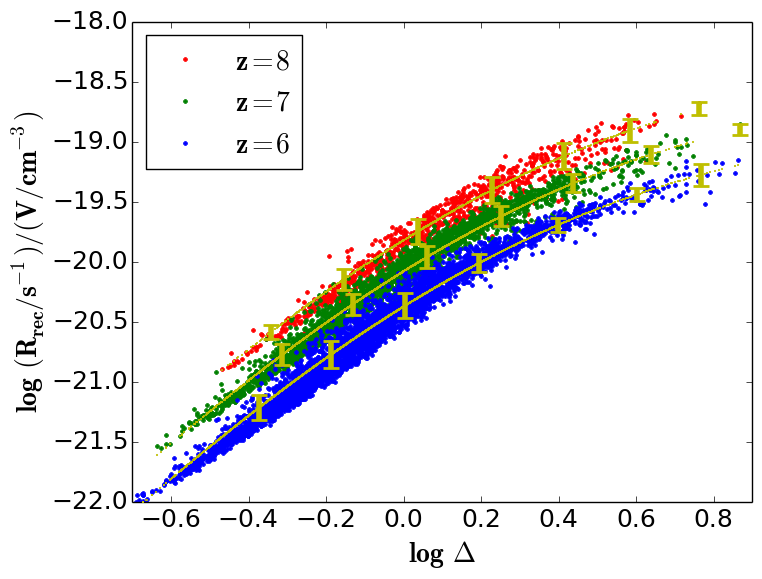}}
\caption{Recombination rate density $\mathrm{\rrec/V}$ computed from {\bf 6/256-RT} simulation at $\mathrm{z = 6}$ (blue-dots), $\mathrm{z = 7}$ (green-dots) and $\mathrm{z = 8}$ (red-dots). Yellow dots show our fitting function, Equation \eqref{rrec}. Vertical error bars represent $1\sigma$ values for ${\rm log\,\Delta}$ bin size of $0.2$. The scatter is fairly small, so is unlikely to systematically impact the 21-power spectrum. It is quite clear that our fitting function fairly reproduces the recombination rate $\rrec$ that is computed from our hydrodynamic simulations.}
\label{fig:gsm}
\end{figure} 

Figure~\ref{fig:gsm} shows the recombination rates per unit volume as
a function of $\Delta$ at $z=6,7,8$.  The yellow vertical
error bars represent the $1\sigma$ scatter in bins of ${\rm log\,\Delta}=0.2$.
At the mean density ($\Delta =0$), the scatter is
$\sigma \sim 0.1$, and is similar for different redshifts and $\Delta$. We
do not expect that this
small scatter will have a significant impact on the derived 21-cm
power spectrum. In general, $\rrec\propto \Delta^2$ as expected since
recombination is a two-body process, though there is a slight flattening
at high-$\Delta$.  Recombination rates are also higher at higher $z$,
since the Universe is denser.

We now determine a fitting function for the recombination rate density, $\mathrm{\rrec/V}$.  We
construct a fitting function $\mathrm{\rrec/V}(\Delta,z)$ as follows:
\begin{equation}\label{rrec}
\mathrm{\frac{\rrec}{V} =  A(1+ z)^{D}  \left[\frac{\left( \Delta/B \right)^{C}}{1+ \left( \Delta/B \right)^{C} } \right]^{4} }\, , 
\end{equation}
and we determine the best-fit values to be 
$\mathrm{A} = 9.85 \times 10^{-24} $cm$^{-3}$s$^{-1}$, $\mathrm{B} = 1.76 $, $\mathrm{C} = 0.82$, 
$\mathrm{D}=5.07$.  Note that the redshift dependence is slightly weaker
than the expected $\rm (1+z)^6$ owing to the evolution of the clumping factor~\citep{fin13}.
Although these fitting values are only applicable for the
cell size that we have chosen, namely $0.375\hmpc$, we find that 
re-binning the hydrodynamical simulation using cells that are half as wide leads to an $\rrec$ that is indistinguishable from our current fit.  Furthermore, re-running our reionization simulations with smaller cells but otherwise the same parameters leads to essentially indistinguishable results.  Hence our results are not sensitive to this choice of cell size.

Equation~\eqref{rrec} thus effectively accounts for the local clumping
factor from the {\bf 6/256-RT} simulation, that we can implement into a
{\sc SimFast21} simulation where such clumping cannot be resolved.  We
can thus compute the recombination rate $\rrec$ in {\sc SimFast21} by
multiplying the recombination rate density $\mathrm{\rrec/V}$ by
the {\sc SimFast21} cell volume.

\subsection{Modifying {\sc SimFast21} to use $\rrec$ and $\rion$}\label{sec:modify}
\begin{table}\centering
 \scalebox{0.8}{\begin{tabular}{ || l || c || c || c ||}\hline
   {\bf Simulation }& {\bf Size ($\hmpc$)} & {\bf No of Cells} & {\bf Resolution ($\hmpc$) }\\ \hline \hline
    {\bf \sc SimFast21} & 210 & $560^{3}$ & 0.375\\ \hline
    {\bf 6/256-RT} & 6 & $16^{3}$ & 0.375\\ \hline
    {\bf 32/512} & 32 & - & -\\ \hline
\end{tabular}}
\caption{ shows a summary of our simulations. We use {\bf 6/256-RT} simulation to computing both $\rrec$ and $\rion$. 
We consider the {\bf 32/512} simulation only to computing $\rion$, and hence we don't divide
the simulation box into cells. }\label{simtab}
\end{table}
To apply the fitting formulae for $\rrec$ and $\rion$
in {\sc SimFast21}, we first smooth the generated density field.
The density field moves relative to the fixed {\sc SimFast21} grid
cells as a consequence of applying the \citet{zeldovich70}
approximation.  We must then smooth the density field onto the grid
cells during the evolution.  Here we implement cloud-in-cell (CIC)
smoothing, where each cell contributes to 8 neighbouring cells.  We
then apply the $\rrec$ fitting formula (eq.~\ref{rrec}) to
the CIC-smoothed density field to generate our recombinations rate
boxes.  For $\rion$, we compute the CIC smoothing directly on the
ionisation field that is generated using the halo catalogs via
Equation~\eqref{eq:nion}.

Given $\rion$ and $\rrec$ computed on the {\sc SimFast21} grid, we can
now use this information to develop a new criterion for whether a
particular region is neutral or ionised.  To do so, we simply compare
the local ionisation rate with recombination
rate, and assign the bubble cells to be ionised if:
\begin{equation}
\centering
\fesc\rion \geq \rrec\ ,
\label{barrier1}
\end{equation}
where $\fesc$ is our assumed escape fraction.  This replaces our
previous criterion based on the efficiency parameter $\zeta$
(eq.~\ref{barrier}).

We note that this criterion assumes that, once a particular region satisfies
this criterion, it is quickly able to ionise the vast majority of
its neutral gas.  This is an approximation, but one expects that
the increasing rate of ionising photon production in the early
Universe together with the dropping cosmic density will in general
yield a fast transition to being fully reionised.  Indeed, such a
quick transition is typically seen for entire simulation
volumes~\citep[e.g.][and we will later show this for our models as
well]{gne00}, and it is physically reasonable to expect such rapid
reionisation will also occur locally.  To rigorously assess its
validity we would need to do a full radiative transfer simulation,
which we leave for future work.

In the original {\sc SimFast21}, an ionised cell was set to
$\mathrm{x_{HII}} = 1$.  In reality, even a fully-reionised patch
of the Universe has some small residual neutral fraction $f_{\rm
resid}$, which depends on the global ionising background and
overdensity~\citep{hui-gnedin97}.  Hence when our ionisation criterion
(eq.~\ref{barrier1}) is satisfied, we set the ionisation
fraction to $\mathrm{x_{HII}} = 1-f_{\rm resid}$, otherwise we leave
it as $\mathrm{x_{HII}} = 0$.

We calculate $f_{\rm resid}$ as follows.  First, we obtain the
\ion{H}{I} photoionisation rate $\Gamma_{\rm HI}$,  which corresponds to the flux of 
ionising photons, from \citet{haa12}.Then, we compute the neutral fraction based on ionisation
equilibrium following \citet{pop09}, which results in
\begin{equation}\label{HIfrac}
f_{\rm resid} = \frac{2C+1-\sqrt{(2C+1)^2-4C^2}}{2C}\, ,
\end{equation}
with
\begin{equation}
C = \frac{n \beta(T)}{\Gamma_{\rm HI}}\, ,
\end{equation}
and where $n$ is the hydrogen number density, $T$ is the gas temperature,
and the recombination rate coefficient $\beta(T)$ function~\citep{ver96} is given by
\begin{equation}
\beta(T) = a\Bigl[ \sqrt{T/T_0}(1+\sqrt{T/T_0})^{1-b}(1+\sqrt{T/T_1})^{1+b}\Bigr]^{-1}.
\end{equation}
For neutral hydrogen, the best-fit parameters are $a=7.982\times 10^{-11}$~cm$^3$s$^{-1}$, 
$b=0.7480$, $T_0=3.148$~K, and $T_1=7.036\times 10^5$~K.   We assume
a temperature of $T=10^{4}$~K at all times, since $f_{\rm resid}$ is only important
at the end of reionisation where the universe becomes optically thin.

At the tail end of reionisation, this results in a non-zero $\mathrm{x_{HI}}$,
unlike in the original code.  We note that this is not a direct prediction of
this simulation, since it scales with the $\Gamma_{\rm HI}$ that we are taking from
\citet{haa12} rather than predicting directly from the model.  In the future we
plan to do this more self-consistently, but since this has no effect for
predicting the 21-cm power spectrum at epochs where reionisation is far from
complete, for the present purposes our current prescription is sufficient.

To summarize, we have modifed {\sc SimFast21} to employ the criterion
specified in Equation~\eqref{barrier1} in order to determine whether
a given region is ionised.  This requires computing the local ionisation
photon rate $\rion$ (eq.~\ref{eq:nion}) and local recombination
rate $\rrec$ (eq.~\ref{rrec}) in each cell, as well as assuming an
escape fraction $\fesc$ that we constrain in the next section.
Ionised cells are set to have a residual neutral fraction
(eq.~\ref{HIfrac}).  Finally, the density field for computing $\rrec$
and ionisation fields are smoothed using a CIC approach.  This
completes the {\sc SimFast21} modifications that we outlined in
\S\ref{sec:simfast21}, and we will investigate how these changes
impact the evolution of reionisation and the 21-cm power spectrum
in \S\ref{sec:impact}.
 
\section{Results for Observables}\label{sec:results}

\subsection{Constraints on the photon escape fraction $\fesc$}\label{sec:fesc}

Our key free parameter is the escape fraction $\fesc$ of ionising
photons from galaxies.  $\fesc$ is essentially set by the amount
of recombinations in the ISM that destroy ionising photons, along
with dust extinction~\citep{kaur15}.  Past large-scale models of
the EoR typically constrain $\fesc$ using observations, particularly
the Thomson optical depth $\tau$ from the CMB
measurements and Ly$\alpha$ observations~\citep{fan06}.
Up until recently, most of these models required high $\fesc$ values
to match the observations.  As an example, the {\bf 6/256-RT} simulation
requires an $\fesc$ that reaches close to unity at high redshifts
and small halo masses.  However, the new $\tau$ values from
\citet{planck15} have eased such constraints, so for instance the
semi-analytic model of \citet{mit15b} finds that they can match
observations with $\fesc\la 20\%$ throughout the EoR.  

An alternative approach to constraining $\fesc$ is to use extremely
high-resolution hydrodynamic simulations to directly predict the
escape of ionising photons from the ISM of primeval galaxies.  Early
simulations by \citet{gne08} predicted $\fesc\sim 1-3\%$.  First
galaxy simulations by \citet{wis14} predicted that $\fesc$ drops
from 50\% to 5\% over halo masses from $10^7-10^{8.5}M_\odot$.
\citet{ma15} used cosmological zoom simulations of more sizable
halos to estimate $\fesc\sim 0.001-0.02$ with an average $\fesc <
5\%$, with no strong dependance on galaxy mass or redshift.  Overall,
it appears that direct predictions of $\fesc$ tend to favor modest
values of the order of a few to ten percent in hydrogen cooling
halos.  While such $\fesc$ values were difficult to reconcile with
previous measurements of $\tau$, \citet{rob15} noted that the new
\citet{planck15} $\tau$ alleviates these tensions.

Here we tune our $\fesc$ to match a suite of EoR observations.  In particular, we
will constrain to observations of the Thomson optical depth ($\tau$)
to the surface of last scattering from \citet{planck15}, ionizing
emissivity density measurements by \citet{bec13}, and the volume-averaged
neutral fraction by \citet{bec15,fan06}.  The measure of $\tau$
primarily constrains the epoch of onset of reionisation, while the
latter two observations primarily constrain the ionising photon
budget at the end of the EoR or shortly after.  Hence matching all
three data sets simultaneously is a significant challenge for
models~\citep[e.g. see][]{fin11,fin15,kuh12,mit11,mit12,mit13}.

\begin{figure}
\setlength{\epsfxsize}{0.5\textwidth}
\centerline{\includegraphics[scale=0.48]{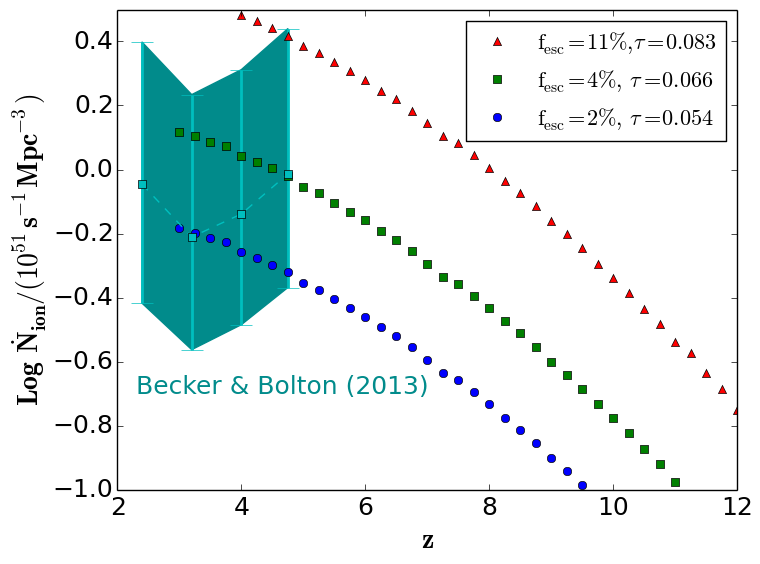}}
\caption{The predicted evolution of the ionizing emissivity density from our fiducial model using the $\rion$ parametrization, Equation\eqref{eq:nion}.  The blue circles, green squares and red triangles represent the ionizing emissivity density of the {\bf Full} model using $\fesc$ = 2\%, 4\%, and 11\% respectively. The darkcyan shaded area comes from \citet{bec13}. It is clearly shown that the three chosen values of $\fesc$ matches the actual, upper and lower limits of the ionizing emissivity by \citet{bec13} as well as the corresponding recent bounds on the reionisation optical depth $\tau$ by \citet{planck15}.}
\label{fig:fesc}
\end{figure} 

We begin by assuming a constant $\fesc$ independent of halo mass
or redshift.  It is important to note that for our {\sc SimFast21} runs, this can be
regarded as the mean escape fraction from all ionising sources
within $0.375\hmpc$ cells which will typically contain a large number of star
forming galaxies; we cannot constrain how $\fesc$
varies for individual galaxies within each cell.  \citet{fin15}
found that, to match the previous WMAP optical depth while still
finishing reionisation by $z\sim 6$ required having a strongly mass-
and redshift-dependent $\fesc$.  However, the larger {\sc SimFast21}
volume and the more recent Planck $\tau$ measurement alters such
requirements.

We perform a {\sc SimFast21} run in a 300~Mpc (comoving) volume
with $560^3$ cells, using our modified version including the
$\rion$ and $\rrec$.  We then vary the $\fesc$ value in order to
match $\tau=0.066$ from Planck.  We are able to match this with
$\fesc=0.04$.  For comparison, changing $\fesc=0.11$ results in a
predicted $\tau=0.083$ which is the $1\sigma$ upper limit from {\it
Planck,} while $\fesc=0.02$ results in a predicted $\tau=0.054$ which is
their $1\sigma$ lower limit.  Hence from this data alone, we constrain
$\fesc=4^{+7}_{-2}$\%. 

We now examine the predicted ionising emissivity, which is the
average rate of ionizing photons emission per unit volume, as
measured from the Ly$\alpha$ forest at redshifts just after
reionisation.  To compute this, we extend our {\sc SimFast21}
simulation to lower-$z$, and we sum the total ionising photon rate
$\rion$ in all cells at a given redshift and divide by the simulation
volume.

Figure~\ref{fig:fesc} shows the predicted values from our simulations as
the red, green, and blue dots for $\fesc=0.11, 0.04, 0.02$, respectively.
The observational range from \citet{bec13} is shown as the cyan region.
It is clear that 4\% provides a good fit to the observations, particularly
at $z\approx 5$ where our simulations are most valid.  
We note that this low value of our $\fesc = 0.04$ has been previously
found using a semi-analytic model by \citet{kul13}. Coincidentally,
the $\fesc$ values to match the upper and lower limits of Planck are
comparable to that required to match the upper and lower limits of the
ionising emissivity measures, showing that these data sets currently
provide comparable constraints on $\fesc$ for our simulations.

\begin{figure}
\setlength{\epsfxsize}{0.5\textwidth}
\centerline{\includegraphics[scale=0.48]{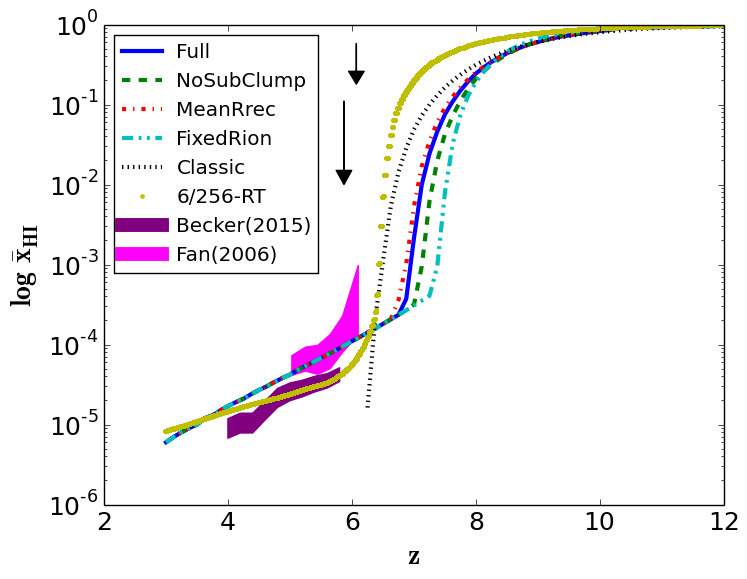}}
\caption{The volume-weighted average neutral fraction, $\mathrm{\bar{x}_{HI}}$, of our models compared to 6/256-RT~\citep{fin15} and observations. Solid blue represents our fiducial model, {\bf Full}. The {\bf NoSubClump} (Dashed, Green), {\bf MeanRrec} (Dash-dotted, Red), {\bf FixedRion} (Dash-dot dotted, Cyan), {\bf Classic} (Dotted, Black) and {\bf 6/256-RT} (Dots, Yellow) are also shown. The shaded magenta and purple show \citet{fan06} and \citet{bec15} measurements respectively.  Vertical arrows represents the recent upper limit constraints by \citet{mcgreer15} at z=6 using Ly$\alpha$ and Ly$\beta$ forests. It is quite clear that adding the residual neutral $f_{\rm resid}$ (eq.~\eqref{HIfrac}) to our fiducial simulations is crucial to match the observation, as opposed to the {\bf Classic} model.  } 
\label{fig:xHI}
\end{figure}

As a final test, we examine the evolution of the cosmic volume-weighted
neutral fraction, as can be probed observationally using the opacity
of the Ly$\alpha$ forest~\citep[e.g.][]{fan06}.  This is shown in
Figure~\ref{fig:xHI}.  Focus for now on the solid blue line labeled
``{\bf Full}", which is the model we are considering here with
$\fesc=0.04$.  In this simulation, the Universe reionises at $z\sim
7$.  This may be slightly higher than that inferred from observations
by \citet{fan06}, but there is some uncertainty on this owing to
the large sightline-to-sightline variation in the mean Ly$\alpha$
opacity~\citep{bec15}.  Hence we find that a 4\% escape fraction
is also broadly consistent with observations of the completion of
reionisation.

\begin{figure*}
\setlength{\epsfxsize}{0.5\textwidth}
\centerline{\includegraphics[scale=0.55]{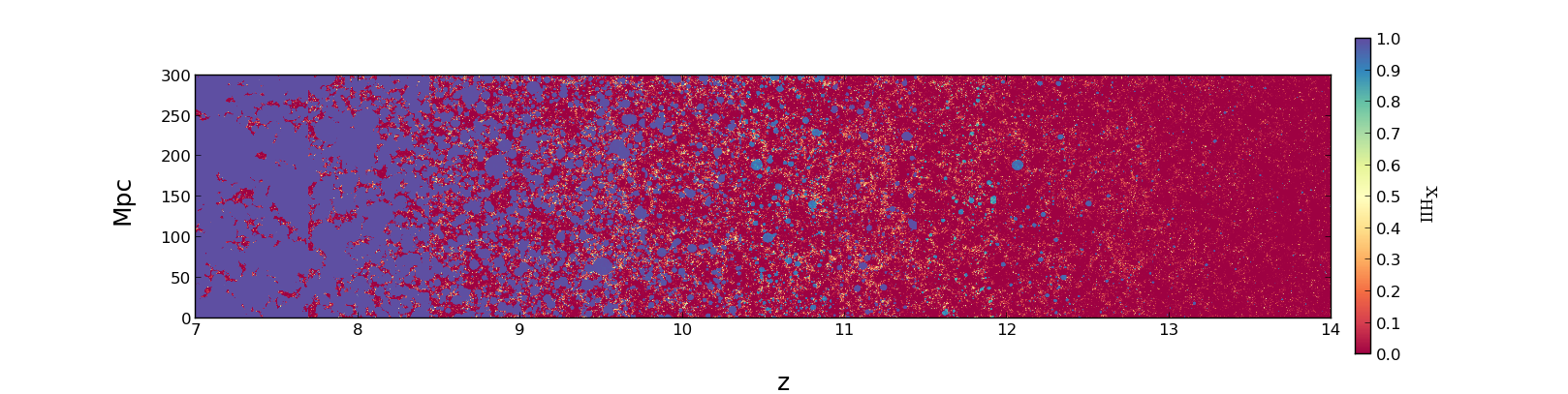}}
\caption{Evolving map of the neutral fraction in our 300 Mpc, $560^3$ cell {\sc SimFast21}
simulation with a 4\% escape fraction.}
\label{fig:ion3}
\end{figure*}

\begin{figure}
\setlength{\epsfxsize}{0.5\textwidth}
\centerline{\includegraphics[scale=0.47]{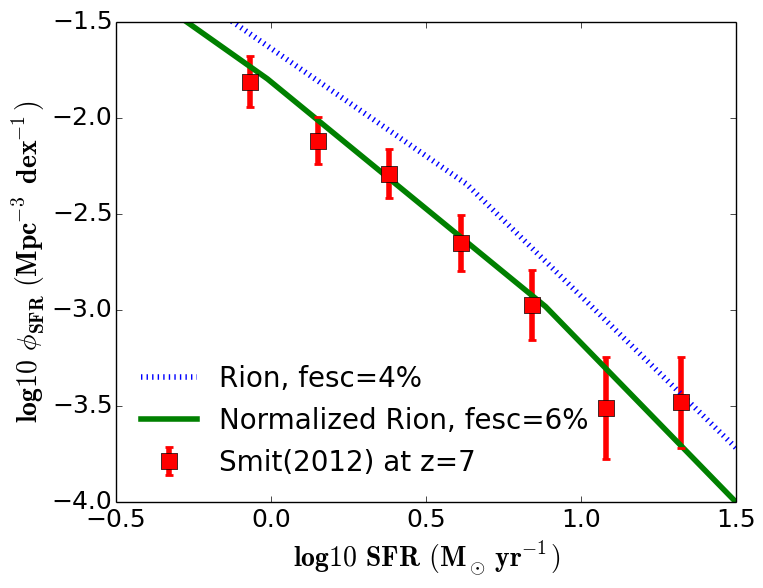}}
\caption{Comparison of the SFR functions from our {\sc SimFast21} using $\rion$ (blue, dotted-line) with the stepwise SFR functions of \citet{smit12} which were derived from the dust-corrected UV Luminosity functions. The SFR at a given halo mass from {\sc SimFast21}, using $\rion$ from hydrodynamic simulations ({\bf 6/256-RT, 32/512}), is higher by a factor of $\times 1.5$  than implied by observations. The green sloid line shows the SRF function from {\sc SimFast21} using the normalized $\rion$ that matches \citet{smit12}.}
\label{fig:sfr}
\end{figure} 

For a more pictorial view, the evolution of the neutral fraction
in our fiducial simulation with $\fesc=0.04$ is depicted in
Figure~\ref{fig:ion3}.  This shows a one-cell-thick ($0.375\hmpc$)
map of the neutral fraction $\mathrm{x_{HI}}$ (Figure~\ref{fig:ion3})
constructed by splicing together the outputs at different redshifts
into a continuous series.  Reionisation begins at $\mathrm{z\sim17}$
and the Universe is reionised by $z\sim 7$. In our model, the EoR is
an extended process since the neutral fraction $\mathrm{x_{HI}}$ drops
slightly from 0.99 to 0.92 as $z\sim 17\rightarrow 11$. It is seen
that the ionised gas forms bubbles of increasing size, corresponding
to a classic inside-out topology where the densest regions are ionised
first\footnote{We note that {\sc SimFast21} essentially assumes such
a topology, since it calculates the ionisation state within bubbles,
and assumes that all the gas within a bubble is fully ionised; it
is not possible to get neutral patches within such bubbles that would
correspond to outside-in reionisation.}. While illustrative,
for actually computing the 21-cm power spectrum we will not use the light
cone in Figure~\ref{fig:ion3} but rather individual snapshots at various
redshifts, since it has been shown by \citet{lap14} that the light cone
effect are small for our box size. The goal of the redshifted 21-cm
EoR observations is to constrain this bubble size distribution and its
evolution through measurements of the power spectrum, thereby providing
constraints on the topology and sources of reionisation.

As noted earlier in Figure~\ref{fig:nion}, it appears
that the global ionisation rate, and by proxy the star formation rate,
at a given halo mass in our hydrodynamical simulations may be high by a factor
of 1.5 $\times$ compared to observations.  To investigate
this in more detail, we show in Figure~\ref{fig:sfr} the star formation
rate function from our {\sc SimFast21} using $\rion$ (blue, dotted-line), compared to observations
by \cite{smit12} at $z=7$.  Indeed, we see that our simulated SFRs are
generally higher by this factor across all galaxies. 

Given that this discrepancy is mostly invariant with star formation
rate, for the purposes of EoR modeling it can be directly translated
into a correction factor on the escape fraction.  In other words,
if we corrected our $\rion$ values down by a factor of 1.5 (normalized
$\rion$) to account for the mismatch in SFRs (green solied line in
Figure~\ref{fig:sfr}), we could raise our $\fesc$ values by a factor
of 1.5 and obtain the exact same results for the 21~cm power spectrum
(see Figure~\ref{fig:pk2}); there is no impact on the predicted
power spectrum of 21-cm fluctuations for a model whose $\fesc$ is
tuned to match observations of the history of reionization.

Hence if we take the observations of \citet{smit12} at face value, we
require an escape fraction of $6^{+10.5}_{-3}$\% to match the various
EoR observations described above.  This would alleviate some of the
discrepancy between our value and that inferred by \citet{rob15},
who argued for $\fesc=20\%$ from observations.  At the same time, it
is not in conflict with direct observational constraints on $\fesc$,
which tend to prefer values of less than $10\%.$

To summarize, our simulation with the new version of {\sc SimFast21}
yields the somewhat remarkable result that a constant escape fraction of
4\%, independent of mass or redshift, is sufficient to match observational
constraints on both the onset and end of reionisation.   This value may
be increased to 6\% when we fine-tune our model to match observed star
formation rate measurements of \citet{smit12} using the normalized $\rion$.  \citet{rob15} also argued
for a relatively modest escape fraction ($ \fesc =20\%$) based on the
new Planck $\tau$, which is somewhat higher than our preferred value.
In contrast, previous works such as \citet{fin15} have argued for a
substantially varying, and generally much higher, $\fesc$.  Even if we
constrain to the WMAP $\tau=0.078$, we still require an escape fraction
of only $\approx 20$\%, still much lower than \citet{fin15}; hence it is
not only just the new value of $\tau$ that is driving our lower $\fesc$.
Additionally, the key aspect is the large volume of the {\sc SimFast21}
simulation that includes more massive halos relative to small-volume,
full-RT simulations.  We will show in \S\ref{sec:massfcn} that, in our
new {\sc SimFast21} model, these moderate-mass halos are an important
contributor to the photon budget, because the ionising photon rate scales
super-linearly with halo mass (eq.~\ref{eq:nion}).  This highlights the
importance of running large-volume simulations, tuned to available
observations, to properly characterise
the escape fractions that are required to complete reionisation.

\subsection{The 21 cm power spectrum}\label{21cmpk}

Using our well-constrained {\sc SimFast21} simulation for the
evolution of the ionisation field, we now make predictions for
our key observable, namely the redshifted 21-cm power spectrum.

Under the assumption that the spin temperature is much higher than
the CMB temperature, we compute the 21-cm brightness temperature
as follows:
\begin{equation}\label{21cm}
\delta T_{b} (\nu) = 23\mathrm{ x_{HI} }\Delta \left( \frac{\Omega_{b} h^{2}}{0.02} \right) \sqrt{  \frac{1+z}{10} \frac{0.15}{\Omega_{m}h^{2}}} \left( \frac{H}{H + d v/dr} \right) \mathrm{mK},
\end{equation}
where $d v/dr$ is the comoving gradient of the line of sight component
of the comoving velocity. Using Equation \eqref{21cm}, we define
the 21-cm power spectrum as follows: $\mathrm{\Delta^{2}_{21} \equiv k^{3}/(2\pi^{2}\, V) < | \delta T_{b} (k)  |^{2}_{k} >}$.

\begin{figure}
\setlength{\epsfxsize}{0.5\textwidth}
\centerline{\includegraphics[scale=0.48]{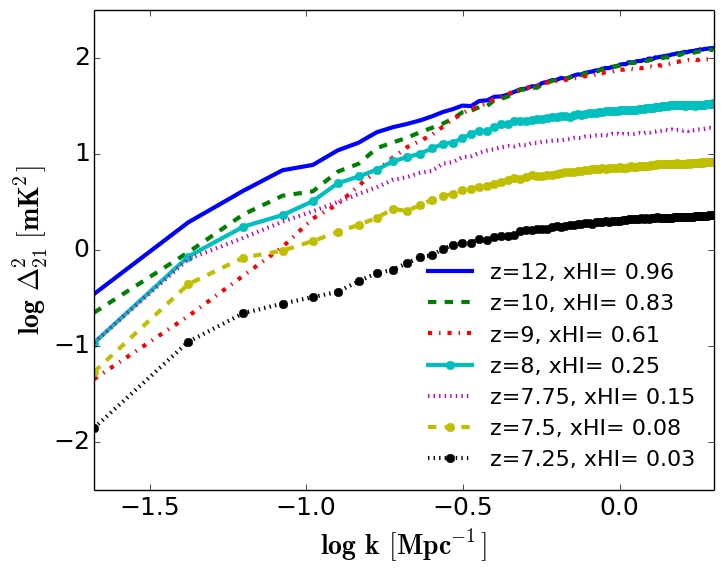}}
\caption{21-cm power spectrum predicted at $\mathrm{z}$=7.25, 7.5, 7.75, 8, 9, 10, 12 from our fiducial 
{\sc SimFast21} run. At early times (${\rm z}$=12), the 21-cm power spectrum traces the density field power spectrum. 
 During $\rm z=12\rightarrow 9$, the presence of more ionised hydrogen in large over-dense regions than under-dense regions suppresses 
 the 21-cm power spectrum. At the intermediate phases, the rapid growth of ionised bubbles boosts the 21-cm power spectrum. At later epochs, 
 when the EoR is nearly complete, the 21-cm power spectrum drops rapidly.}
\label{fig:21cmevol}
\end{figure}

\begin{figure*}
\setlength{\epsfxsize}{0.5\textwidth}
\centerline{\includegraphics[scale=0.3]{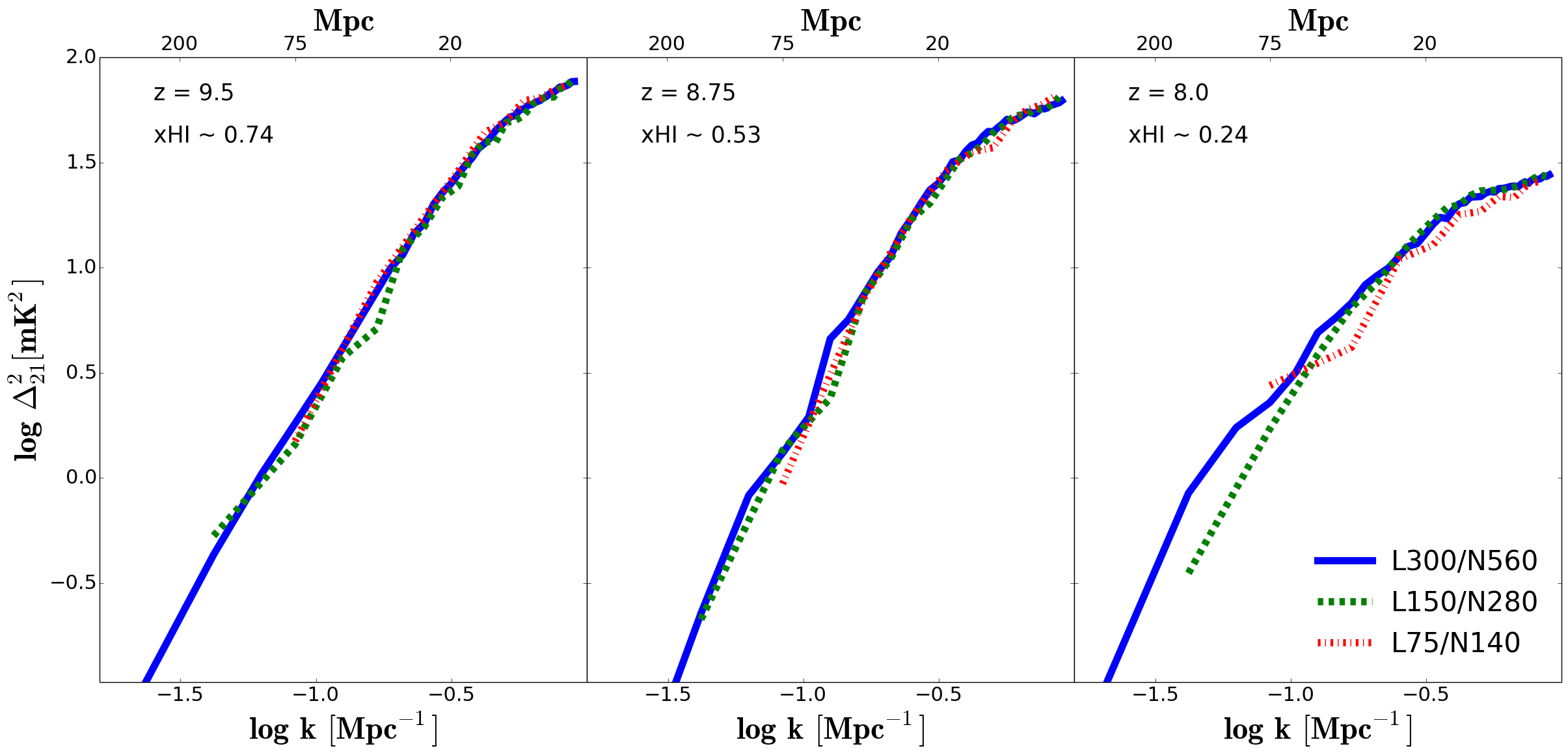}}
\caption{Volume convergence of the 21-cm power spectrum for
{\sc SimFast21} runs with a box size of 300~Mpc (blue, solid), 150~Mpc (green, dashed),
and 75~Mpc (red, dashed-dotted).  The convergence at all redshifts is excellent up to about
one-quarter of the box size.}
\label{fig:21cmconv}
\end{figure*}
Figure~\ref{fig:21cmevol} shows the 21cm power spectrum
at $\mathrm{z}$=7.25, 7.5, 7.75, 8, 9, 10, 12 from our 300~Mpc,
$560^3$-cell {\sc SimFast21} simulation.  The evolution in
Figure~\ref{fig:21cmevol} qualitatively follows the pattern described
in ~\citet{lidz08}: At early times (z=12), the 21cm power spectrum
traces the density power spectrum, because virtually all of the gas
is neutral.  During the interval $z=12\rightarrow 9$, the tendency
for reionisation to begin in the largest overdensities offsets the
tendency for such overdensities to host high concentrations of
neutral gas. As soon as the most overdense regions are ionized, 
the density field of \emph{neutral} hydrogen is overall much more 
homogeneous than prior to reionisation, suppressing large-scale power 
in 21cm fluctuations.
At later epochs, the the ionised regions become larger, leading to
more large-scale power.  In essence, this behavior is reflective
of inside-out reionisation where the inhomogeneities are first
prominent on smaller scales and then move to larger scales.

As emphasized by \citet{sob14}, a spatially-inhomogeneous recombination
rate suppresses fluctuations on scales larger than $\sim0.1$
Mpc$^{-1}$ (this effect is also visible in Figure 1 of ~\citet{lidz08},
although they did not emphasise it). The major advance in our work
with respect to \citet{lidz08} and \citet{sob14} and others is a demonstration
that including a superlinear dependence of ionising efficiency on
halo mass amplifies large-scale fluctuations, partially restoring
the flatness of the 21cm power spectrum; we will return to this
point in our discussion of Figure~\ref{fig:pk2}.

Figure~\ref{fig:21cmevol} represent our predictions for the evolution
of the 21-cm power spectrum from the EoR.  In a follow-up paper,
we will examine the detectability of $\Delta^2_{21}$ for ongoing
and upcoming 21-cm EoR experiments.  For this work, we focus on
studying how $\Delta^2_{21}$ is impacted by the physical modeling
variations that we have implemented into {\sc SimFast21}.

\subsection{Numerical convergence}

Our chosen simulation volume is generally limited by our computational
capabilities, together with the requirement that our cell size match
the chosen cell size over which we have computed our recombination
rate in the {\bf 16/256-RT} simulation.  Here we check whether our results
are robust to our choice of volume by running simulations with
smaller volumes (keeping the cell sized fixed).  This will also
allow us to empirically determine the largest robustly predicted
scale in our simulation for a given box size.

Figure~\ref{fig:21cmconv} shows the 21-cm power spectrum $\Delta_{\rm
21cm}(\rm k)$ calculated from boxes with length 150~Mpc (green,
dashed) and 75~Mpc (red, dot-solid), along with our fiducial 300~Mpc
box (blue solid).  These show  $\Delta_{\rm 21cm}(\rm k)$ over
scales from twice the cell size up to the full box size.  The
corresponding physical scale $2\pi/k$ is shown along the top axis.  Three panels
show $\rm z=9.5, 8.75, 8$ which correspond to global neutral fractions
of roughly three-quarters, one-half, and one-quarter, respectively.

Generally, the numerical convergence with box size is very good,
particularly during the early phases of reionisation.  At $\rm z=8$,
some deviations are seen at the large scales (small $k$), which
start at about one-quarter of the box size or larger, but they are
typically less than a factor of 2 in the power.  We conclude that
our simulations can robustly predict the 21-cm power spectrum over
the range of scales from a few times the cell size up to one-quarter
of the box size, over the redshift range where there is a significant
global neutral fraction and hence 21-cm signal.

\section{Impact of varying ionisations and recombination assumptions}\label{sec:impact}

Our main improvement from the previous version of {\sc SimFast21}
is a more physically-motivated characterisation
for the ionising source population and small-scale recombinations.
Here we quantitatively investigate the impact of these new
parameterizations for $\rion$ and $\rrec$ on the reionisation history
and morphology, in comparison with previous assumptions.  We do so
by essentially reverting our new code back towards the original
code one piece of physics at a time, so that we can isolate the
impact of each new physical component.

To do this, we run five simulations with a box size L$=300$ Mpc and
N$=560^{3}$ on a side using the same density field and halo catalogs
(i.e. the same cosmology), only with different astrophysical
assumptions for $\rion$ and $\rrec$.  Because this changes the
evolution of the ionisation field, in order to make the comparison
more uniform, we re-tune the photon escape fraction $\fesc$, or in
the case of the original {\sc SimFast21} the efficiency parameter $\zeta$,
in order to achieve the Planck $\tau=0.066\pm0.016$.  By doing so, we can compare
these models at the same redshift and neutral fraction more
meaningfully.

The five simulations are as follows:
\begin{itemize}
\renewcommand{\labelitemi}{\tiny$\blacksquare$}
\item {\bf Full:} This is our fiducial model in which we use our new parameterizations $\rion$ and $\rrec$ in Equation~\eqref{barrier1} to identify the ionised regions, assuming $\fesc = 4\%$. 
\item {\bf NoSubClump:}  Similar to {\bf Full}, but using inhomogeneous recombinations computed from local cell's densities (=$\alpha_{a}\mathrm{n_{H}^{2}}$) with no contribution from sub-grid clumping, we re-tune to obtain $\fesc = 2.5\%$.
\item {\bf MeanRrec:}  Similar to {\bf Full}, but using a spatially-homogeneous recombination rate computed from the mean hydrogen cosmological density at each $z$; $\fesc=1.5\%$.  
\item {\bf FixedRion:} Similar to {\bf Full}, only using Fixed Rion per halo mass = $9\times 10^{49} \mathrm{sec^{-1}}$ which corresponds to the value for $\mathrm{M_{h} = 10^{8} M_{\odot}}$ and assuming $\fesc = 100\%$. 
\item {\bf Classic:} This run with the original {\sc SimFast21} using Equation\eqref{barrier} to identify the ionised regions with $\zeta =11$.
\end{itemize}

Our {\bf Full} model is what we have used to make the predictions
presented so far.  The {\bf NoSubClump} and {\bf MeanRrec} simulations
do not use the $\rrec$ taken from our {\bf 16/256-RT} box, but rather
compute recombinations locally on a cell-by-cell basis, and globally
using the cosmic mean density, respectively.  Variations among these
three runs can thus be used to isolate the scale at which recombinations
are important.  The {\bf FixedRion} simulation is analogous to using a
constant efficiency parameter, but still uses our {\bf Full} recombination
model; hence comparing to the {\bf Full} model can be used to assess
the importance of employing a halo mass-dependent ionisation rate.
The {\bf Classic} case is the original {\sc SimFast21} code, which uses a
constant efficiency parameter and no explicit recombinations.

We now focus on the variations among these models for the key observables that
we have described before, and provide a physical interpretation for
the differences that we see.

\subsection{Global neutral fraction history}

\begin{figure*}
 \centerline{ \includegraphics[width=18cm,height=12cm]{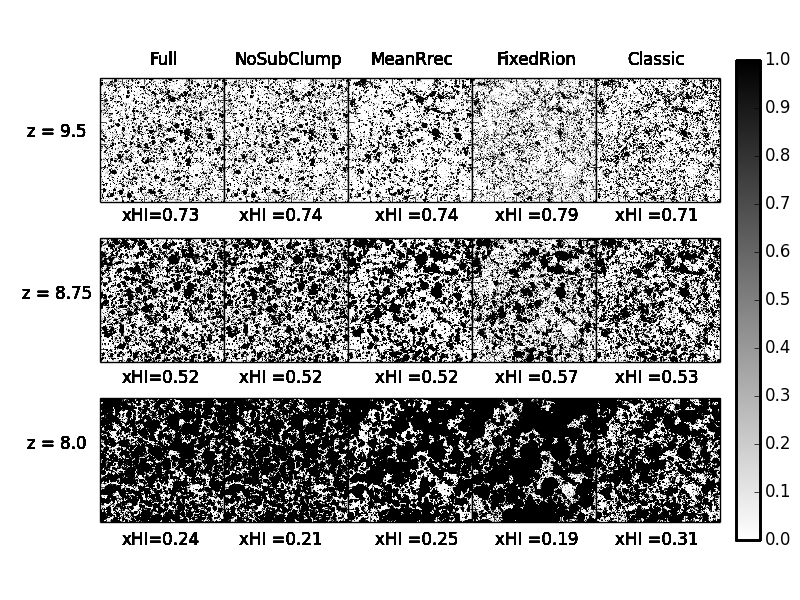}}
  \caption{The ionisation maps of the five models at z=9.5 (top-row), z=8.75 (middle-row) and z=8.0 (bottom-row). White regions are neutral whereas Black regions are ionized. It is clear that the {\bf Full} and {\bf NoSubClump} models display similar morphologies across all EoR phases, suggesting that the local sub-clumping effects ($\leq 0.5 {\rm Mpc}$)  have no significant contribution to the EoR on large scales ($\geq 100 {\rm Mpc}$). The {\bf FixedRion}, {\bf MeanRrec} and {\bf Classic} models display similar morphologies only with different bubble sizes due to the variation in the physical assumptions.} 
  \label{fig:maps}
\end{figure*}

Figure \ref{fig:xHI} shows the volume-weighted neutral fraction
($\mathrm{\bar{x}_{HI}}$) evolution produced by our five models,
as labeled.  We also show this evolution taken directly from the
{\bf 6/256-RT} simulation as the yellow dots.  Observations by
\citet{fan06} and \citet{bec15}, with $1\sigma$ range, are indicated
by the shaded regions. Vertical arrows represents the recent ``model-independent"
upper limits by \citet{mcgreer15} at $z=6$ using the Ly$\alpha$ and Ly$\beta$ forests.

At high-$\rm z$, the onset of reionisation is similar for all models,
primarily because we have tuned them to match the {\it Planck} $\tau$
value which is most directly a constraint on the onset of reionisation.
Comparing our {\bf Full} simulation first to the {\bf 6/256-RT} simulation,
we note that while the shape of $\mathrm{\bar{x}_{HI}(z)}$ is similar,
reionisation occurs earlier by $\rm \Delta z\sim 0.5$.  As mentioned
before, this owes to the small dynamic range of {\bf 6/256-RT} that fails
to capture: (i) the very earliest stages driven by the rarest over densities
corresponding to the longest-wavelength fluctuations. (ii) the larger halos that are 
important particularly during the later stages of the EoR.

Comparing the {\bf Full} to {\bf Classic} runs, it is seen that the
{\bf Classic} run reionises the Universe later, more like $\rm z\sim
6.5$.  Since this model has no explicit model for recombinations, the early ionisations
are very effective, which reduces the amount of ionisations needed
at early times in order to reproduce the {\it Planck} $\tau$ value.
This results in less ionising photons overall, which delays
reionisation.  The strength of this effect is best quantified by
comparing to the {\bf FixedRion} run, which like the {\bf Classic}
case has a constant ionising photon efficiency per unit halo mass.
Comparing these we can see that ignoring recombinations can shift
the end of the EoR by $\rm \Delta z\sim 1$.

The impact of recombinations is seen by comparing the {\bf Full},
{\bf NoSubClump}, and {\bf MeanRrec} models.  The differences in
the ionisation history are relatively minor, with {\bf NoSubClump}
producing slightly earlier reionisation.  In this comparison, {\bf
Full} and {\bf MeanRrec} produce very similar global reionisation
histories, but we will later see they differ significantly in terms
of topology.

Overall, changing the nature of the ionisation sources has a larger
impact, at fixed $\tau$, than varying the recombination methodology.
Nonetheless, constraining to match $\tau$ results in models having
an end of reionisation all within $\rm \Delta z\sim 1$ of each other.
Hence the global evolution of the neutral fraction is relatively
insensitive to our modifications to {\sc SimFast21}, with the largest
difference being relative to a model with no recombinations
({\bf Classic}) or a small-volume hydrodynamic simulation that does
not yield large galaxies ({\bf 6/256-RT}).

\subsection{Ionizations Maps}\label{sec:maps}

We now explore how the topology of reionisation varies amongst our
five different physical models.  This is relevant to the 21-cm power
spectrum since it focuses on topological features such as the
distribution of ionised bubble sizes that directly reflects in the
power spectrum.  Furthermore, 
forthcoming observation with the SKA can in principle directly map the neutral
gas distribution. To get a flavour for the sorts of topology variations
introduced by varying our input physics, we begin by examining maps
of the ionisation field evolution.

Figure \ref{fig:maps} shows maps of the ionisation field for the
five models across different phases of the EoR at $z=9.5, 8.75,
8.0$ (top to bottom).  Overall, the Universe becomes more ionised
at later epochs, and the global neutral fractions are generally
similar at the same $z$.  Nonetheless, there are clear differences in the topology
of the ionised regions amongst the various models.

The {\bf Full} and {\bf NoSubClump} models display quite similar
morphologies across all EoR phases as seen in their ionisations
maps (first two columns). This implies that including the sub-grid
clumping effects through $\rrec$ does not strongly affect the EoR
topology, and suggests that computing recombinations using only the
local cell's densities (as in {\bf NoSubClump}) is sufficient to
properly model the EoR.  This is tantamount to saying that the local
clumping factor within each cell is close to unity, and the main
variations in the clumping factor occur on scales larger than our
cell size; indeed, we have checked that this is the case within the
{\bf 6/256-RT} run.  While this suggests that including $\rrec$ was perhaps
superfluous, this outcome was not obvious from the outset. 
This perhaps conflicts with the conclusions of \citet{raic11}; their Figure~2 
suggests that the local clumping fluctuations are quite important.
We then expect that the {\bf Full} and {\bf NoSubClump} models will accordingly
produce the same ionisation and 21-cm power spectra.

The {\bf MeanRrec} ionisation maps show relatively larger ionised
bubbles (third column) than in our fiducial model.  Hence ignoring
the density fluctuations and clumping on large scales and assuming
recombinations based on the cosmic mean density is a poor approximation.
This occurs because the {\bf MeanRrec} model has fewer recombinations,
because the recombinations primarily occur in the dense regions,
and the recombination rate scales roughly as the square of the
density.  Therefore, the ionising photons can propagate farther and
create larger bubbles.

At early stages of the EoR (at z=9.5), the {\bf FixedRion} model
(fourth column) produces smaller ionised bubbles than other models.
This occurs because of the interplay of recombinations and the
ionising source locations.  Given that our reionisation topology
is generally inside-out (dense regions ionising first), this means
that the dense regions with the largest halos live in regions that
are most rapidly recombining.  In the {\bf Full} model, the increasing
strength of ionising photon output with halo mass helps to offset
these rapid recombinations, and power ionising fronts out of the
dense regions.  However, in the {\bf FixedRion} model the large
halos have lower ionising output, and hence the resultant bubbles
grow less before stalling.  Conversely, in low-density regions
with smaller halos, {\bf FixedRion} will produce larger bubbles.
Hence overall, the bubble size distribution is more uniform in
this case, which will be evident in the power spectra we consider
below.

It is interesting to note that the {\bf MeanRrec} and {\bf FixedRion}
models share similar morphologies at the intermediate phase of the
EoR ($z=8.75$), which implies that inhomogeneous $\rrec$ has roughly
a similar level of an effect as including mass-dependent ionisation.
Given that we have constrained all models to match the {\it Planck}
optical depth, a ``crossover" at $\mathrm{x_{HI}}\sim 0.5$ is perhaps not
surprising.

Finally, the {\bf Classic} (fifth column) displays a morphology
that is between {\bf FixedRion} and {\bf MeanRrec} morphologies.
Its form of ionisation is most similar to {\bf FixedRion}, but it
includes recombinations implicitly through the constrained efficiency
parameter $\zeta$ (eq.~\ref{barrier}) as a direct suppression of
the number of ionising photons output in each cell.  Effectively,
this cell-based suppression ends up being stronger than the
recombinations in {\bf MeanRrec} and weaker than that in {\bf
FixedRion}, which sets the {\bf Classic} morphology between those
of {\bf MeanRrec} and {\bf FixedRion}

These results show that assumptions regarding the ionisation and
recombination rates in large-scale models can have a significant effect on
the topology of reionisation and its evolution.  

\subsection{Ionized mass fraction as a function of density}

\begin{figure*}
\setlength{\epsfxsize}{0.5\textwidth}
\centerline{\includegraphics[scale=0.3]{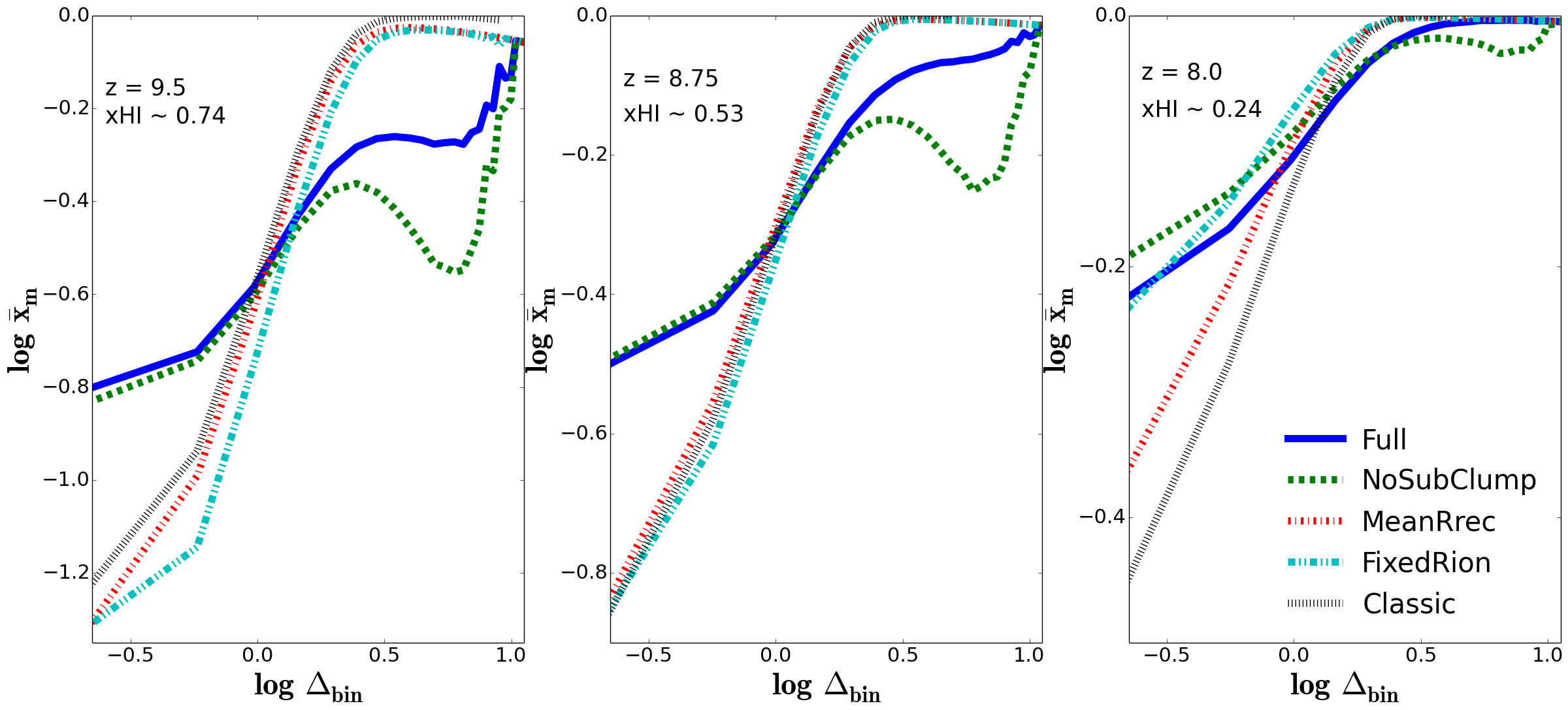}}
\caption{The mass-weighted ionized fraction $\mathrm{\log \bar{x}_{m}}$ evolution of the five models for given overdensity bin $\mathrm{\log \Delta_{bin}}$. \textsc{Left:} Early EoR phase, \textsc{Middle:} Intermediate EoR phase and \text{Right:} Final EoR phase.  The over-dense regions ionize first while the under-dense regions take longer to become ionised, which shows that the EoR proceeds in an inside-out fashion.}
\label{fig:dbin}
\end{figure*}
\begin{figure*}
\setlength{\epsfxsize}{0.4\textwidth}
\centerline{\includegraphics[scale=0.3]{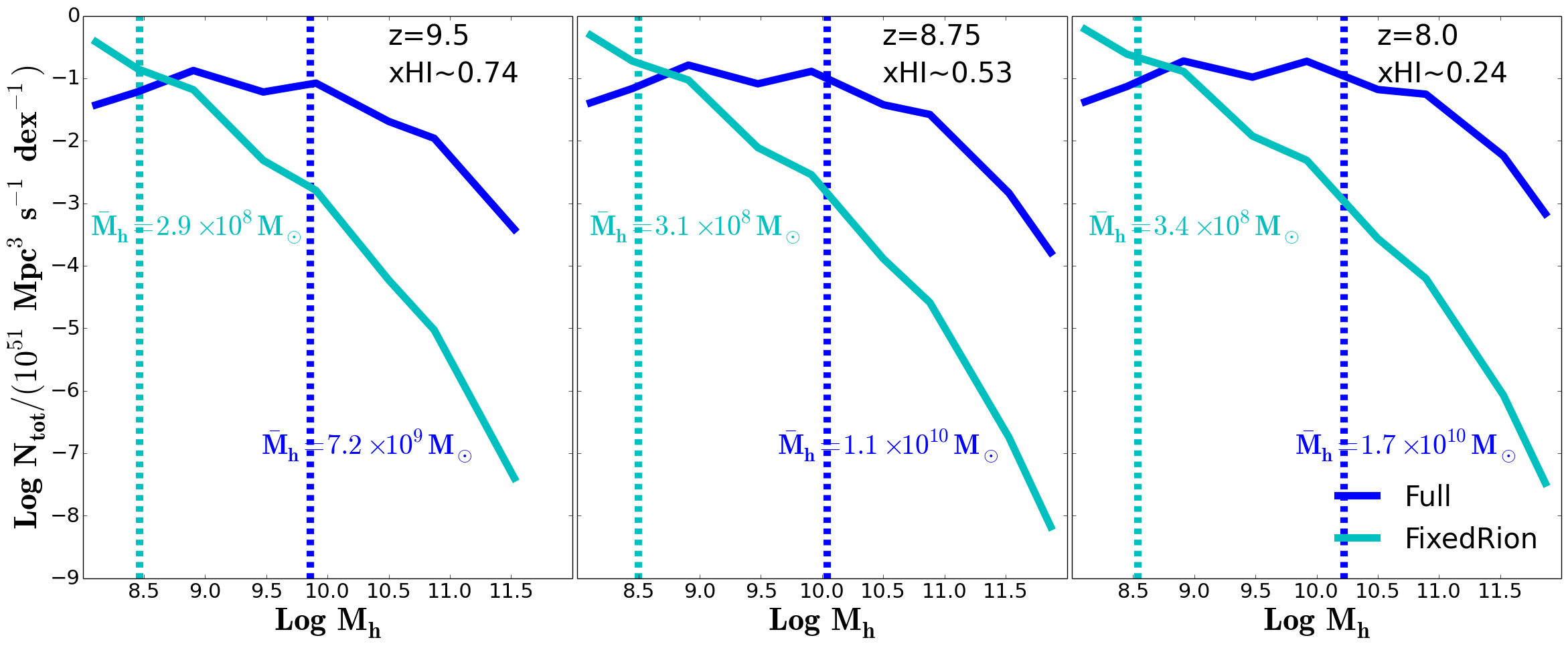}}
\caption{Total ionizing emissivity ($\mathrm{N_{tot}}$) of our fiducial model (blue, solid) and the {\bf FixedRion} (cyan, solid)  for halo mass bin size of 0.5. Vertical dashed lines represent the ionisation-weighted halo mass ($\mathrm{\bar{M}_{h}}$) of the {\bf Full} (blue) and {\bf FixedRion} (cyan) models. It is evident that the dominant halo mass during EoR is much larger in our fiducial model than in the {\bf FixedRion} model by about 1-2 order of magnitude.}
\label{fig:hist}
\end{figure*}

We now quantify the topology shown in the maps in the previous
section.  The approach we use here is to consider the ionised mass
fraction as a function of local overdensity, to better understand
how our physical assumptions are impacting the ionisation state of
the gas in various environments.

Figure \ref{fig:dbin} shows the mass-weighted global ionised fraction
$\rm \bar{x}_m$ as a function of density $\mathrm{\Delta_{bin}}$ of our
five models at three different redshifts $z=9.5, 8.75, 8.0$ (left
to right), when the neutral fraction is approximately $0.75, 0.5,$
and $0.25$, with some variations between the models (the
values quoted on the figure are the average $\mathrm{x_{HI}}$ of all models at each $\rm z$).  

In all models and at all times, the over-dense regions
($\log \mathrm{\Delta_{bin}} > 0.0$), where the halos and hence the
photon sources are more numerous, are more ionised while low-density
regions such as voids ($\log \mathrm{\Delta_{bin}} < 0.0$ ) take longer
to become ionised. This is due to the gradual progress of the
ionisations fronts through the surrounding regions regions into low
density regions~\citep{ian06}, and the relatively low clumping
factor~\citep{fin13} and recombination rate particularly in the
late EoR, which results in these high-density regions remaining ionised.
Hence in all these runs, as we saw in the {\bf Full} case earlier
(Figure~\ref{fig:ion3}), reionisation proceeds in an inside-out
fashion.

We first compare {\bf Full} (solid blue) with {\bf NoSubClump}
(dashed green).  In the moderate and low-density regions, these
models are essentially identical, except at late times when the
ionised fraction in {\bf NoSubClump} is slightly higher (likely an
artifact of tuning to the same $\tau$ value).  Sub-cell gas clumping
thus only has an effect in relatively more overdense regions, which
physically makes sense since such clumping occurs on small scales.
The impact of sub-cell clumping is maximized at overdensities $\mathrm{\Delta_{bin}} \sim 5-10$,
whereas in the most dense regions the ionisations overwhelm
recombinations regardless of the clumping.  However, the trend is
opposite to one might naively expect:  The {\bf NoSubClump} model
actually has a lower ionisation fraction, indicating more recombinations,
than the {\bf Full} model.  This happens because the {\bf NoSubClump}
model computes the reionisation rate assuming a fully ionised cell
at the cell's density, but the $\rrec$ fitting formula used in {\bf
Full} includes the effect that dense regions have more neutral gas,
and hence the recombination rate is lowered compared to the fully
ionised case.

Comparing {\bf MeanRrec} to {\bf Full}, we see much more dramatic
differences than compared to {\bf NoSubClump}.  As we have seen
from the maps earlier, assuming recombinations only at the mean
cosmic density produces quite a different topology.  The ionised
fraction is significantly higher in the high-density regions, and
lower in the low-density regions, with a crossover around the mean
density ($\Delta=1$).  Since recombinations scale as the square of
the density, dense regions have lower recombination rates in {\bf
MeanRec} leading to more ionised gas, and conversely low-density
regions have higher recombination rates leading to less ionised
gas.

The {\bf MeanRrec}, {\bf FixedRion} and {\bf Classic} have roughly
the comparable behaviour of $\mathrm{\bar{x}_{m}}$ during the early
phases of the EoR. This also was clear in their topologies (Figure
\ref{fig:maps}).  Once these models are constrained to have matching
$\tau$ values which governs their early EoR evolution, using a mean
density recombination rate with inhomogeneous sources turns out to
be roughly equivalent in topology to using a fixed ionising output
per unit mass and inhomogeneous recombinations.  The {\bf Classic}
model falls in between these two extremes, with fixed ionising
efficiency and the recombinations implicitly tied to the ionisations.
At lower redshifts, the three models start to diverge at low
densities, as the higher recombination rates at $\log\Delta<0$ in the
{\bf MeanRec} model lead to lower ionised fractions than in the
{\bf FixedRion} model.  The {\bf Classic} model is still lower,
since the recombinations are tied to ionisations which are very low
in the low-density regime.  Interestingly, in this regime
the {\bf FixedRion} model becomes similar to the {\bf Full} case.

Overall, our results show that including both inhomogeneous
recombinations and ionisations is important to produce the correct
topology evolution of the EoR.  While neglecting sub-cell recombinations has
little impact, assuming either recombinations that are based on the
mean density or tied to the ionisation rate results in dramatically
different ionised gas topologies.  Likewise, using a fixed ionising
photon output per unit halo mass results in a substantially different
topology during the neutral-dominated phase of the EoR, though it
becomes more similar to the fiducial case at later stages.

\subsection{Ionizing photon output versus halo masses}\label{sec:massfcn}

An interesting quantity for observing the EoR is the
mass scale of galaxies providing the bulk of the ionising photons.
This is important for future observational programs with for instance
the {\it James Webb Space Telescope}, which aims to directly detect
these galaxies.  Given that we track the ionising output of halos,
we can straightforwardly determine the ionising photon distribution as a
function of halo mass, and assess the impact of our new input physics
on this.  We consider only on our fiducial ({\bf Full}) and
{\bf FixedRion} models, to compare the total amount of emitted
ionising photons per halo mass in each model at fixed $\tau$, 
since we are not concerned with recombinations here.

Figure \ref{fig:hist} shows histograms of the total ionising
emissivity $\mathrm{N_{tot}}$ for halo mass bins of the {\bf Full}
(blue) and {\bf FixedRion} (cyan) at $z=9.5,8.75,8$ (from left to
right).  The {\bf FixedRion} model, with an assumed constant escape
fraction and constant ionising output per unit halo mass, results
in a steeply declining ionising emissivity with $M_h$ that mimics
the steep halo mass function.  In contrast, the {\bf Full} model
has greater ionising output per $M_h$ for more massive halos,
shifting the distribution towards higher halo mass.  

The vertical dashed lines (blue:{\bf Full}, cyan:{\bf FixedRion})
in Figure~\ref{fig:hist} represent the ionisation-weighted halo mass
${\bar{M}_{h}}$ of all halos at the corresponding redshifts
(z=9.5,8.75,8.0). ${\bar{M}_{h}}$ thus represents the halo
mass limit above (or below) which 50$\%$ of the total ionizing
photons is being emitted. We see that the ${\bar{M}_{h}}$
is $\sim{10^{10}M_{\odot}}$ in our fiducial model whereas
it is $\sim{10^{8.5}M_{\odot}}$ in {\bf FixedRion}, with only
a mild dependence on redshift.

Hence a model assuming a fixed ionising output per unit halo mass
would predict that reionisation is dominated by extremely small
galaxies near the hydrogen cooling limit, while a more realistic
model for ionising photon output suggests that moderate-mass halos
are responsible for reionisation.  The latter scenario bodes well
for future direct observations of galaxies driving reionisation,
since galaxies in $\sim 10^{10}M_\odot$ halos are likely to be
detectable with {\it Webb}.

\subsection{Ionizations and 21-cm power spectra}

\begin{figure*}
\setlength{\epsfxsize}{0.6\textwidth}
\centerline{\includegraphics[scale=0.3]{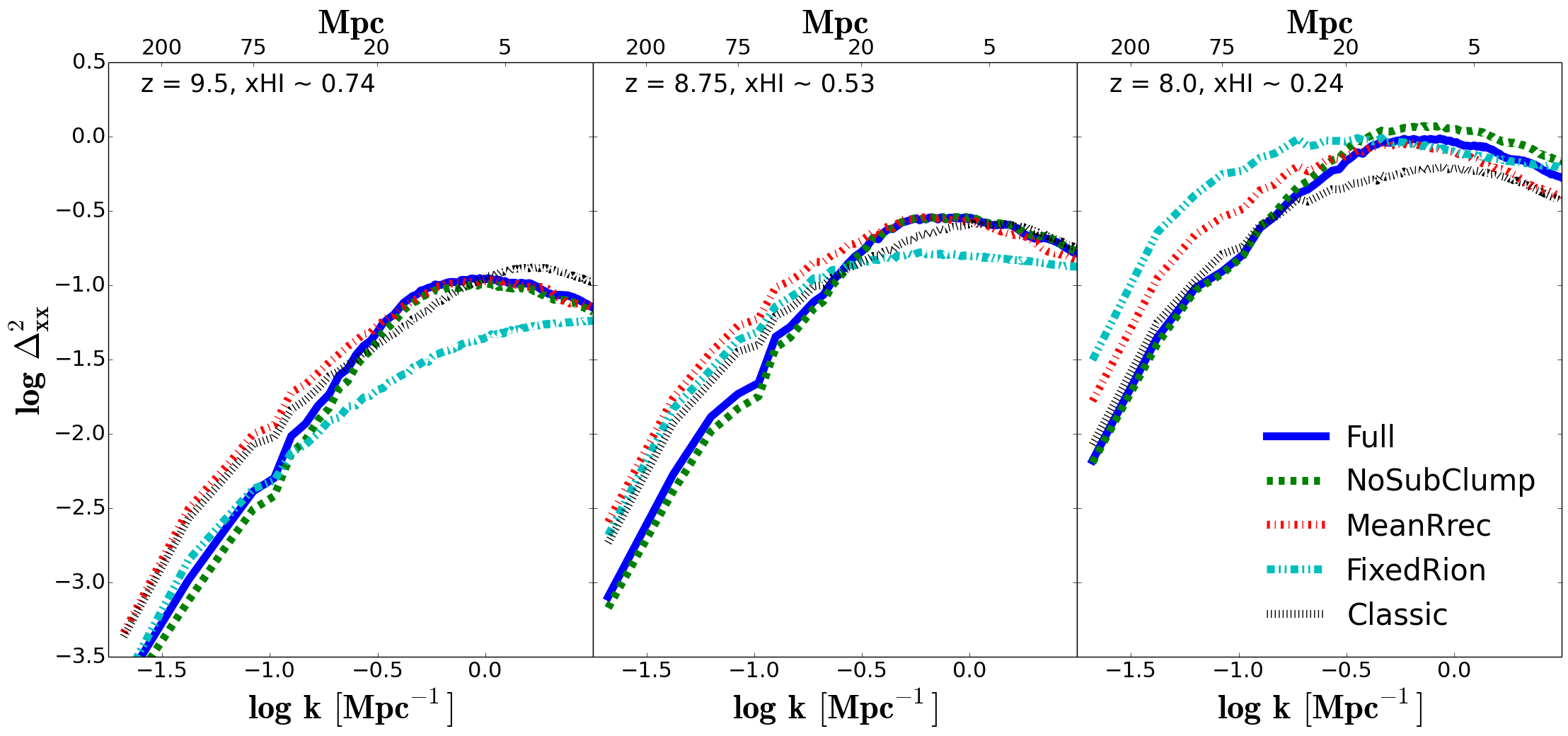}}
\caption{The power spectra of the ionisation fields  of the five models at different stages during the EoR.   Variations in physical assumptions can result in $\times 2-3$ variations in the ionisation power spectrum such as: $\rrec$ suppresses the large-scale power spectrum while $\rion$ boosts the small-scale power spectrum. }
\label{fig:pk1}
\end{figure*}

\begin{figure*}
\setlength{\epsfxsize}{0.6\textwidth}
\centerline{\includegraphics[scale=0.3]{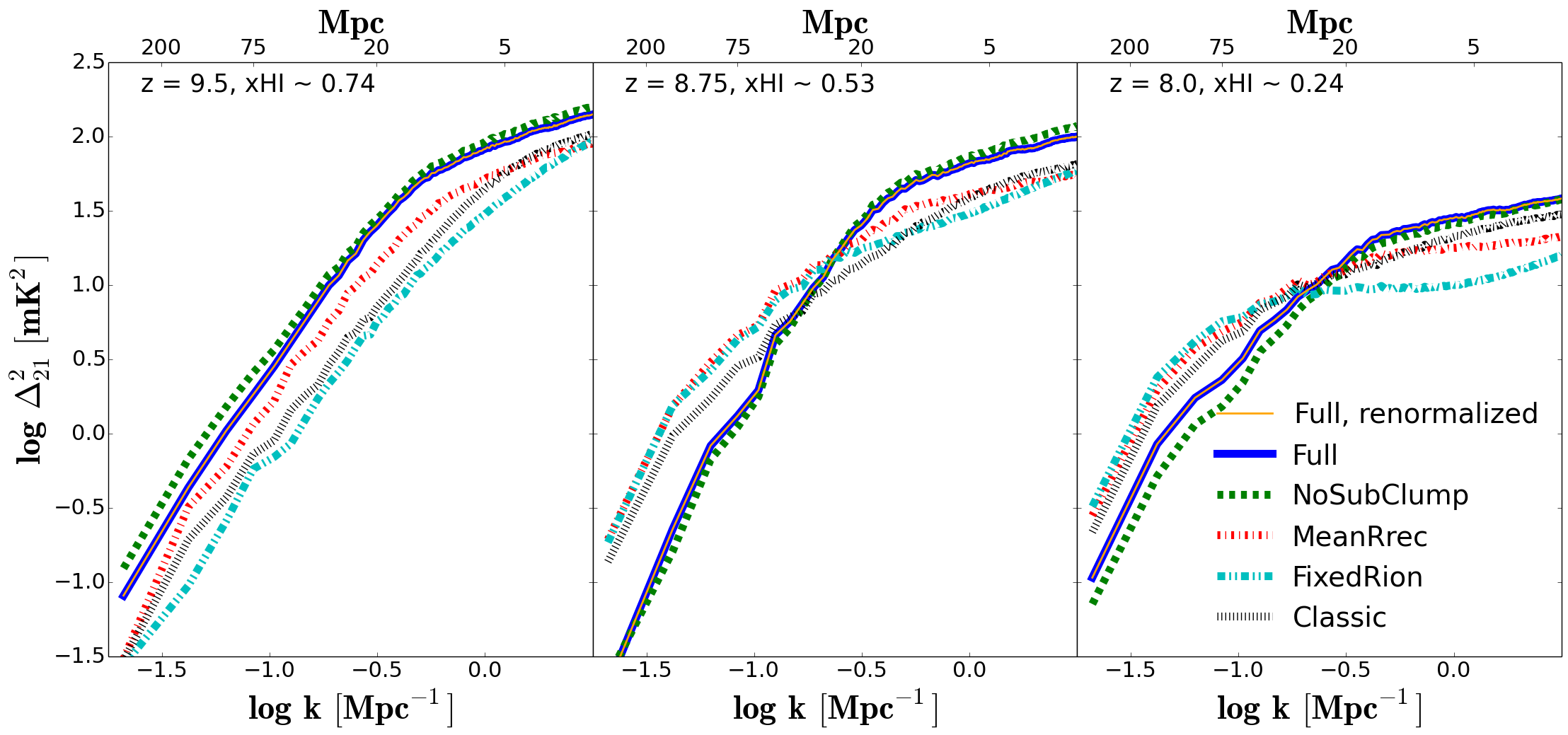}}
\caption{The power spectra of the 21-cm signal of the five models at different stages during the EoR.  Variations in physical assumptions can result in $\times 2-3$ variations in the 21-cm power spectrum such as: $\rrec$ suppresses the large-scale power spectrum while $\rion$ boosts the small-scale power spectrum.}
\label{fig:pk2}
\end{figure*}

We now examine how these physical variations impact the ionisation
and 21-cm power spectra.  The 21-cm power spectrum is the key
observable that will be obtained in the forthcoming 21-cm EoR
observations.  We compute the ionisation power spectrum as
$\mathrm{\Delta^{2}_{xx} \equiv k^{3}/|(2\pi^{2}\, V) < | x_{HII}|^{2}
>/x^{2}_{HI}}$, analogous to the 21-cm power spectrum that was
introduced in \S\ref{21cmpk}.

In Figure \ref{fig:pk1} and \ref{fig:pk2}, we compare the ionization
fields and 21-cm  power spectra of our fiducial model ({\bf Full})
to other models at different phases of the EoR, respectively.
We begin by comparing the 21-cm power spectra from our
fiducial model ({\bf Full}) using the actual $\rion$ (blue, solid),
that requires $\fesc=4\%$ to match the observations excluding SFR
function measurements, with the renormalized $\rion$ (orange, solid)
that requires $\fesc=6\%$ to match the SFR functions measurements.
We see, as expected, that the 21-cm power spectra with and without
the renormalization to $\rion$ are identical, which confirms that
the normalization of $\rion$ can be directly compensated by tuning
our free parameter $\fesc$ to higher or lower values as implied by
observations.

\begin{figure*}
\setlength{\epsfxsize}{\textwidth}
\centerline{\includegraphics[scale=0.3]{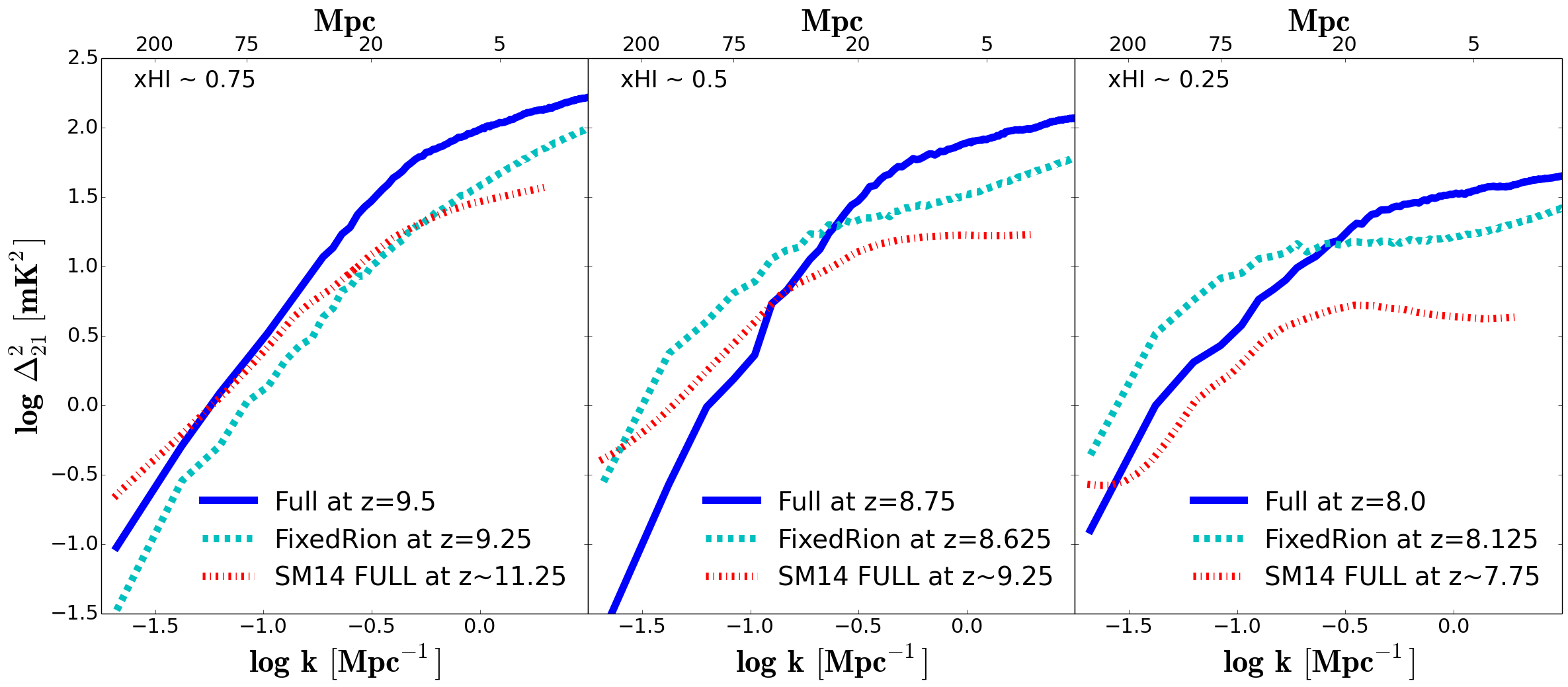}}
\caption{Comparison of the predicted 21-cm power spectrum between our {\bf Full} (blue, solid) and {\bf FixedRion} (dashed, cyan) models with the {\bf FULL} model (dash-dotted, red) of SM14 \citep{sob14} at different neutral fraction.  It is evident that the shape of the 21-cm power spectrum predicted from  {\bf FixedRion} model is similar to that of SM14 {\bf FULL} model at all epochs, since both models share similar physical assumptions. Our {\bf Full} model  produces more power on small scales due to introducing non-linearly mass-dependent ionisations via $\rion$. }
\label{fig:com_pk}
\end{figure*}

As with the topology, the {\bf Full}
and {\bf NoSubClump} models produce approximately similar power
spectra of the ionisation fields and 21-cm signal through all EoR
phases.  The differences in topology at moderate overdensities are
sufficiently rare to not impact the overall power spectrum appreciably.
This confirms that the sub-clumping effects have little significance
to the EoR 21-cm observables.

To further test the effects of sub-grid clumping, we
also ran the {\bf NoSubClump} model with the same $\fesc =4\%$ that
is used in the {\bf Full} model, without matching to the same $\tau$.
We confirmed that the {\bf NoSubClump} model produces mostly identical
results at the same neutral fraction for this slightly different
$\fesc$. This confirms our finding, that the sub-clumping has no
significant contribution to the EoR process, is robust and independent
of the way we used to do this comparison with the {\bf Full} model.

As we have seen in studying the overall topology, the {\bf Full}
and {\bf NoSubClump} models are very similar, while the {\bf
MeanRrec}, {\bf FixedRion}, and {\bf Classic} cases show similar
evolutionary trends.  Not fully accounting for both non-linear
ionisations and inhomogeneous recombinations tends to raise the ionisation field
power spectrum on large scales relative to smaller scales, which
results in a lower 21-cm power spectrum by a factor of a few.  Hence,
for instance, compared to the old version of {\sc SimFast21} during the
early stages of the EoR, our new code predicts almost $4\times$ as
much power on large scales, and twice as much on small scales.

The shape of the power spectra continue to be different at later
stages, at which time the {\bf Full} model ends up with less large
scale power but more small-scale power, as visually evident from
the ionisation maps in Figure~\ref{fig:maps}.  The physical reasons
were discussed in \S\ref{sec:maps}, but basically arise because
properly accounting for recombinations tends to slow the ionisation
fronts when compared to {\bf MeanRrec}, while the {\bf FixedRion}
produces relatively more photons in low-density regions from low-mass
halos which causes more large-scale bubbles.  The {\bf MeanRrec},
{\bf FixedRion}, and {\bf Classic} runs themselves show different evolutionary
trends, with {\bf FixedRion} increasing its ionisation and 21-cm
power spectra faster than the other two.

A comparison of the {\bf MeanRrec} and {\bf Classic}
models in Figure~\ref{fig:pk2} illustrates the significance of
including a realistic treatment for the dependence of ionising
efficiency on halo mass.  Both models effectively adopt a
spatially-homogeneous recombination rate and are explicitly tuned
to reproduce $\tau$.  However, the {\bf MeanRrec} (like the {\bf
Full} model) adopts the ionising efficiency in Figure~\ref{fig:nion}
and Equation~\eqref{eq:nion}, which concentrates ionisations in
overdense regions.  Consequently, the {\bf MeanRrec} model predicts
more large-scale power at all times than the {\bf Classic} model.
This effect counteracts the suppression of large-scale power
owing to spatially-inhomogeneous recombinations~\citep{sob14}.
Note that the adopted dependence in Equation~\eqref{eq:nion} is not
arbitrary -- it is taken from high-resolution hydrodynamic
simulations that have been shown to be consistent with a range of
observations of galaxies, absorbers, and reionisation~\citep{fin15}.
Our results indicate that it is an important factor to consider
when predicting the 21-cm power spectrum.

 Finally, we compare the predicted 21-cm power spectrum
from our new version of {\sc SimFast21} with those predictions from
another semi-numerical simulation \citep[their {\bf FULL} model]{sob14}
in Figure~\ref{fig:com_pk}. SM14 {\bf FULL} simulation accounts for
inhomogeneous recombinations but uses only homogeneous ionisations
via a constant efficiency parameter ($\zeta$, similar to our {\bf
Classic} model).  Thus, the SM14 model shares similar physical
assumptions to our {\bf FixedRion} model.  From Figure~\ref{fig:com_pk},
it is clear to see that the shape of the 21-cm power spectrum
predicted from {\bf FixedRion} model is qualitatively similar to
that of the SM14 model at all epochs, but there is a tendency for
the {\bf FixedRion} model to produce somewhat more power as the EoR
proceeds; this is possibly related to different density field
contribution from different z. However, it is quite clear that our
{\bf Full} model predicts more power on small scales than models
with homogeneous ionisations owing to non-linear ionisations.

To summarize, our new prescription for tracking non-linear
ionisations and inhomogeneous recombinations results in predictions that differ
significantly from models that do not include these effects, in
particular yielding less power on scales of $\ga 10$~Mpc, but more
power on smaller scales.  Consistent with \citet{sob14}, we find
that introducing inhomogeneous recombinations suppresses the power
spectra of the ionisation field and 21-cm signal on large scales. We also find that
introducing non-linear ionisations (via $\rion$) boosts the small scale power
spectra of the ionisation fields and the 21-cm signal.  These results
highlight the importance of carefully considering the details of 
ionising sources and recombinations in making accurate predictions
for future 21 cm EoR studies.

\section{Conclusion}\label{sec:conc}

We have predicted the 21-cm power spectrum during the
epoch of reionisation from a new and improved version of our
semi-numerical code {\sc SimFast21}. This new version has been
modified to incorporate halo mass-dependent ionisation rates ($\rion$)
and density-dependent recombination rates ($\rrec$) rates as a
function of redshift.  We parameterise $\rion$ and $\rrec$ from
small-volume, high-resolution radiative hydrodynamic simulation
({\bf 6/256-RT},~\citealt{fin15}) and a larger cosmological hydrodynamic
simulation ({\bf 32/512},~\citealt{dav13}), that incoporate galaxy formation
physics that has been well-constrained to match a wide range of
observations down to lower redshifts.  We have implemented these
scalings into {\sc SimFast21} to identify the ionised regions, as
opposed to using a uniform ionising efficiency parameter per halo
mass ($\zeta$) and no explicit recombinations as in the old version
of {\sc SimFast21}.  Using this, we have studied the evolution of
the neutral fraction and 21-cm power spectrum during reionisation.

Our main key findings are as follows:
\begin{itemize}

\item The mass-dependent ionisation rate $\rion$ scales super-linearly
with halo mass as ${M_{h}^{1.41}}$, which is consistent with
the SFR-${M_{h}}$ relation that previously found by
\citet{fin11}.  The recombination rate $\rrec$ scale roughly as the density
squared, though with deviations at high overdensities.  Both display
fairly tight relations that can be well-captured by analytic
fitting formulae (eq.~\ref{eq:nion} and eq.~\ref{rrec}).

\item We tune our one free parameter, the ionising photon escape
fraction, to be $\fesc=4^{+7}_{-2}\%$, in order to simultaneously
match three key EoR observables: (i) The optical depth to Thomson
scattering from \citet{planck15}; (ii) the ionising emissivity
measured at $z\sim 5$ from \citet{bec13}; (iii) the neutral fraction
near end of reionisation from \citet{fan06,bec15,mcgreer15}.  This
low $\fesc$, independent of halo mass and redshift, highlights the
importance of running large-volume simulations to properly characterise
the escape fraction $\fesc$ and hence the ionising photon budget
to complete reionisation.  Note that this $\fesc$ represents the
escape fraction averaged over all halos within a 0.375$\hmpc$
(comoving) cells, rather than $\fesc$ from individual galaxies.
To further concurrently match the SFR function
measurements by \citet{smit12}, a renormalization of $\rion$ by a
factor of 1.5 is required, which then implies an escape fraction
of 6\%.  This renormalization of $\rion$ and $\fesc$ has no impact
on the 21-cm power spectrum or other observables, since the total
photon output remains the same.

\item During the early EoR, the 21-cm power spectrum drops on large
scales while staying constant on small scales, as small ionisation
bubbles counteract the overall drop in cosmic \ion{H}{i} density.
At later stages, the bubbles grow larger and the power on large
scales recovers.  After the global neutral fraction drops below
$\sim 10\%$ (at $z\sim 7.5$ in our simulation), the 21-cm power
spectrum drops rapidly.

\item Reionisation occurs earlier in our {\sc SimFast21} run than
in the 6/256-RT simulation by $\mathrm{\Delta z \sim 0.5}$, due to
the small dynamic range of 6/256-RT that fails to capture the very
earliest stages driven by the rarest overdensities, as well as large
halos that are important during the later stages of the EoR.  These
results are well-converged with respect to simulation volume up to
scales $\sim 1/4$ of the box size.

\item Introducing non-linearly mass-dependent ionisations ($\rion$): (i)  increases
the duration of reionisation; (ii) boosts the 21-cm power at all
scales by $\times 2-3$ during the early EoR; (iii) boosts the
small-scale 21-cm power by $\times 2-3$ while lowering the
large scale ($\ga 5 \hmpc$) power during the late EoR.  Qualitatively
similar trends hold true for the ionisation field power spectra
({\bf Full} versus {\bf FixedRion}).

\item Including spatially homogeneous recombinations using a globally-averaged 
recombination rate results in significantly more power on large scales
and less power on small scales, since ionisation bubbles are able
to grow to larger scales ({\bf Full} versus {\bf MeanRrec}).

\item Including clumping effects on scales below our cell size
($0.375\hmpc$) does not significantly affect the overall EoR topology
or 21-cm power spectrum ({\bf Full} versus {\bf NoSubClump}), though
it does result in a lower ionisation fraction at moderate overdensities
($\Delta \sim 5-10$).

\item In agreement with \citet{sob14}, we find that inhomogeneous
recombinations ($\rrec$) matter a great deal ({\bf Full} or {\bf
NoSubClump} versus {\bf MeanRec}) in suppressing the large-scale
power spectrum, but accounting further for the
detailed dependence of recombination rate on density at scales
smaller than our cell size changes results minimally, and only at
fairly high overdensities ({\bf Full} versus {\bf NoSubClump}).

\item Compared to the previous version of {\sc SimFast21} ({\bf
Classic}), our new version produces more small-scale 21-cm power
and less large-scale power.  It is generally most similar to a model
that assumes a globally-averaged recombination rate ({\bf MeanRrec})
or assumes a constant ionising output per unit halo mass ({\bf
FixedRion}).

\item The power spectrum from the model of \citet{sob14} that account
for recombinations but retain a fixed ionising output per unit halo
mass is qualitatively similar to our {\bf FixedRion} case, albeit
with generally a lower amplitude.

\item Our {\bf Full} model shows a significantly higher median halo
mass for ionizing photon output of $\sim 10^{10} M_{\odot}$, as
opposed to $\sim 10^{8.5} M_\odot$ in the case of a constant ionising
efficiency.  This suggests that the majority of galaxies responsible
for reionisation may be detectable with {\it Webb}; we leave a more
detailed examination of this for future work.

\item Incorporating the non-linear mass-dependent ionisation causes
reionisation to complete at a later epoch by $\Delta z \sim
1$; this is a larger impact than that obtained by varying the
recombination methodology.

\end{itemize}

Taken together, our results suggest that the details of how one
models ionisation and recombinations can impact the strength and
shape of the 21-cm power spectrum up to a factor of $\sim \times
3$.  This is smaller than current observational uncertainties which
are dominated by systematics, but might be 
significant for upcoming facilities such as HERA and SKA-Low.  This work represents a step
forward in accurately modeling the physical processes occuring during 
reionisation on large scales.  In the future we will use these models
to make more detailed predictions for the observability of the 21-cm
power spectrum at various observationally-accessible epochs and scales.

We note that we have not considered any potential exotic sources
of reionising photons.  For instance, a recent paper by \citet{mad15}
argues for quasars contributing significantly more than previously
thought, indeed perhaps driving reionisation.  Such a population
of accreting black holes would provide an additional source of
ionising photons that would once again vary how reionisation proceeds.
Likewise, our radiation-hydrodynamic simulation did not properly
track the contribution of mini-halos (i.e. those below the H cooling
limit) to the ionising photon budget during reionisation, but there
are no observational constraints that limit this.  Our {\sc
SimFast21}-based modeling framework provides a way to explore these
variations and their impact on observables such as the 21-cm power
spectrum, in order to facilitate more optimal scientific interpretation
of forthcoming 21-cm and other EoR observations.

 \section*{acknowledgements}
The authors acknowledge helpful discussions with Andrei Mesinger,
Sourav Mitra, Suman Majumdar, Marta Silva and Daniel Cunnama.  SH is supported by
the Deutscher Akademischer Austauschdienst (DAAD) Foundation.  RD
and SH are supported by the South African Research Chairs Initiative
and the South African National Research Foundation. MGS acknowledges
support from the South African Square Kilometre Array Project, the
South African National Research Foundation (grant 92788) and FCT
under grant PTDC/FIS-AST/2194/2012. This work was also supported
by NASA grant NNX12AH86G.  Part of this work was conducted at the
Aspen Center for Physics, which is supported by National Science
Foundation grant PHY-1066293.  Computations were performed at the
cluster ``Baltasar-Sete-Soi'', supported by the DyBHo-256667 ERC
Starting Grant, and the University of the Western Cape's ``Pumbaa"
cluster.

\end{document}